\documentclass[prd,12pt,showpacs,floatfix]{revtex4}
\usepackage{graphicx}
\usepackage{latexsym}
\usepackage{amsmath}
\newcommand{\bq}{\begin{equation}}
\newcommand{\eq}{\end{equation}}
\newcommand{\bqn}{\begin{eqnarray}}
\newcommand{\eqn}{\end{eqnarray}}

\newcommand{\lb}{\label}
\begin{document}
\title{Stable and ``bounded excursion"  gravastars, and black holes in Einstein's theory of gravity} 
\author{P. Rocha $^{1,3}$}
\email{pedrosennarocha@gmail.com}
\author{R. Chan $^{2}$}
\email{chan@on.br}
\author{M.F.A. da Silva $^{1}$}
\email{mfasnic@gmail.com}
\author{Anzhong Wang $^{1,4}$}
\email{anzhong_wang@baylor.edu}
\affiliation{$^{1}$ Departamento de F\'{\i}sica Te\'orica, 
Instituto de F\'{\i}sica, Universidade do Estado do Rio de Janeiro, 
Rua S\~ao Francisco Xavier 524, Maracan\~a
20550-900, Rio de Janeiro - RJ, Brasil\\
$^{2}$ Coordena\c{c}\~ao de Astronomia e Astrof\'{\i}sica, Observat\'orio
Nacional, Rua General Jos\'e Cristino, 77, S\~ao Crist\'ov\~ao  20921-400, Rio
de Janeiro, RJ, Brazil\\
$^{3}$ Universidade Est\'acio de S\'a, Rio de Janeiro, RJ, Brazil\\
$^{4}$ GCAP-CASPER, Department of Physics,
Baylor University, Waco, TX 76798, USA}

\begin{abstract}
 
Dynamical models of prototype gravastars are constructed and studied. The 
models are the  Visser-Wiltshire three-layer gravastars, in which  an 
infinitely thin spherical shell of a perfect  fluid with the equation
of state $p = (1-\gamma)\sigma$ divides the whole spacetime into two regions, 
where the internal region is  de Sitter, and  the external is  Schwarzschild. 
When $\gamma < 1$ and $\Lambda \not= 0$, it is found that in some cases the  
models represent stable 
gravastars, and in some cases they represent ``bounded excursion"  stable 
gravastars, where the thin shell is oscillating between two finite radii, 
while in some other cases they collapse until the formation of black holes. 
However, when $\gamma \ge 1$, even with $\Lambda \not= 0$, only black holes 
are found. In the phase space, 
the region for both stable gravastars and  ``bounded excursion"  gravastars 
is very small in comparison to that of black holes, although it is not 
completely empty.  

\end{abstract}
\preprint{arXiv: xxxxxxxx}
\maketitle

\section{Introduction}

\renewcommand{\theequation}{1.\arabic{equation}}
\setcounter{equation}{0}

As alternatives to black holes,  gravastars have received some attention
recently \cite{grava}, partially due to the tight connection between the 
cosmological constant and a currently accelerating universe \cite{DEs}, 
although very strict observational constraints on the existence of such 
stars may exist \cite{BN07}. In the original model of Mazur and  Mottola (MM) 
\cite{MM01}, gravastars consist of five layers:
an internal core $0 < r < r_{1}$, described by the de Sitter universe, 
an intermediate thin layer of stiff fluid $r_{1} < r < r_{2}$, and an external 
region $ r > r_{2}$, described by the Schwarzschild solution,
\bq
\lb{1.1}
ds^{2} = -f(r) dt^{2} + \frac{1}{f(r)} dr^{2} 
   + r^{2}\left(d\theta^{2} + \sin^{2}\theta d\varphi^{2}\right),
\eq
where the function $f(r)$ is given by $f(r) =  1 - {2{\cal{M}}}/{r}$,
in the units where $c = 1 = G$. In addition, in such a setup,  two infinitely thin shells
also  appear, respectively, on the hypersurfaces $r = r_{1}$ and $r = r_{2}$. By properly choosing
the free parameters involved, one can show that the two shells can have only tensions but with
opposite signs \cite{MM01}. Visser and Wiltshire (VW) argued that such five-layer models can be
simplified to three-layer ones \cite{VW04}, in which the two infinitely thin shells and the 
intermediate region are replaced by one infinitely thin shell, so that the function $f(r)$ 
in the metric (\ref{1.1}) is 
given by
\bq
\lb{1.3}
f(r) = \begin{cases} 1 - \frac{2{\cal{M}}}{r}, &  r > a(\tau),\\
1 - \left(\frac{r}{l}\right)^{2}, &  r < a(\tau),
\end{cases}
\eq
where $r = a(\tau)$ is a timelike hypersurface, at which the infinitely thin shell is located,
and $\tau$ denotes the proper time of the thin shell. The constant $l \equiv \sqrt{3/\Lambda}$ 
denotes  the de Sitter radius.
On the hypersurface $r = a(\tau)$ Israel junction conditions yield
\bq
\lb{1.4}
\frac{1}{2}\dot{a}^{2} + V(a) =  0,
\eq
where an overdot denotes the derivative with respect to the proper time $\tau$
of the thin shell. %, and   $E$ is a constant. 
Therefore, in the region $ r > a(\tau)$ the spacetime is locally Schwarzschild, while in the
region $ r < a(\tau)$ it is locally de Sitter.  These two different regions are connected through
a dynamical infinitely thin  shell located at $r = a(\tau)$ to form a new spacetime of gravastar.

Two different types of stable gravastars are identified by VW, stable gravastars and
 ``bounded excursion" gravastars.

{\bf Stable gravastars}: In this case,  there must exist a radius $a_{0}$ such that
\bq
\lb{1.5}
V\left(a_{0}\right) = 0, \;\;\; V'\left(a_{0}\right) = 0, \;\;\;
V''\left(a_{0}\right) > 0,
\eq
where a prime denotes the ordinary differentiation with respect to the indicated argument.
If and only if there exists such an $a_{0}$ for which the above conditions are satisfied,
the model is said to be stable. 
Among other things, VW found that there are many equations of state for which the gravastar
configurations are stable, while others are not \cite{VW04}. Carter studied  the same
problem and found new equations of state for which the gravastar is stable \cite{Carter05}, 
while De Benedictis {\em et al} \cite{DeB06} and Chirenti and Rezzolla \cite{CR07} 
investigated the stability of the original model
of  Mazur and  Mottola against axial-perturbations, and found that gravastars are stable to
these perturbations. Chirenti and Rezzolla also showed that their quasi-normal modes differ from
those of a black hole of the same mass, and thus can be used to discern a gravastar from a
black hole. Early work on dynamical thin shells with de Sitter interior and (A)dS-RN exterior
can be found in \cite{ALT99}.

{\bf ``Bounded excursion" gravastars}: As VW noticed, there is a less stringent notion of 
stability, the so-called ``bounded excursion" models, in which there exist two radii $a_{1}$ 
and $a_{2}$ such that
\bq
\lb{1.6}
V\left(a_{1}\right) = 0, \;\;\; V'\left(a_{1}\right) \le 0, \;\;\;
V\left(a_{2}\right) = 0, \;\;\; V'\left(a_{2}\right) \ge 0,
\eq
with $V(a) < 0$ for $a \in \left(a_{1}, a_{2}\right)$, where $a_{2} > a_{1}$. 

Lately, we studied
this type of gravastars \cite{JCAP}, and found that, among other things,  such configurations
can indeed be constructed, although   the region for the formation 
of this type of gravastars is very small in comparison to that of black holes.

In this paper, we generalize our previous work \cite{JCAP} to the case where the equation of state 
of the infinitely thin shell is given by $p= (1-\gamma)  \sigma$ with $\gamma$ being a constant. 
When $\gamma = 0$ it reduces to the case  we studied in \cite{JCAP}. We shall first construct 
three-layer dynamical models, in analogy to the VW models, and then show both  stable gravastars
of the both types and black holes exist for $\gamma < 1$ and $\Lambda \not= 0$. However, when 
$\gamma \ge 1$ even  with $\Lambda \not= 0$, only black holes are found. In the phase space,  
the region of gravastars and the one of black holes  are non-zero, although the former is much 
smaller than the latter. The rest of 
the paper is  organized as follows: In Sec. II we shall study various cases, in which
all the possibilities of forming black holes, gravastars, de Sitter, and Minkowski spacetime
exist. In Sec. III we present our main conclusions.

\section{Formation of Gravastars and Black Holes from Gravitational Collapse of  Prototype
Gravastars}

\renewcommand{\theequation}{2.\arabic{equation}}
\setcounter{equation}{0}

For spacetimes given by,
\bq
\lb{2.1c}
ds^2= \begin{cases} c_{-}\left(-c_{-}f_{-} dv_{-} + 2 dr_{-}\right)dv_{-} + r_{-}^2 d^{2}\Omega, 
&  r < a(\tau),\\
c_{+}\left(-c_{+}f_{+} dv_{+} + 2 dr_{+}\right)dv_{+} + r_{+}^2 d^{2}\Omega, 
&  r > a(\tau), 
\end{cases}
\eq
Lake found that the Israel's junction conditions yield  two independent equations \cite{LAKE},
\bqn
\lb{2.1a}
& & \dot{a}^{2} = (\frac{a}{2M})^{2}(f_{+}-f_{-})^{2}-\frac{1}{2}(f_{+}+f_{-})+(\frac{M}{2a})^{2}, \\
\lb{2.1d}
& & \dot{M}+ 8 \pi a \dot{a} p = 4 \pi a^{2} [T_{\alpha \beta}u^{\alpha}n^{\beta}],
\eqn
where  $M \equiv 4 \pi a^{2} \sigma$,
$\sigma$ denotes the energy density of the shell, and $p$ its pressure.
In this work, we shall consider the case where inside the shell the spacetime is de Sitter, 
and outside it is Schwarzschild, namely
\bq
\lb{2.1e}
f_{+}(r)  =  1 - \frac{2{\cal{M}}}{r},\;\;\; 
f_{-}(r) = 1 - \left(\frac{r}{l}\right)^{2}, 
\eq
as that studied  in ~\cite{JCAP}. The only difference is the equation of state of the thin shell, 
where in the present paper we consider a more general case in which it is taken as,
\bq
\lb{2.1f}
p=(1- \gamma)  \sigma,
\eq
with $\gamma$ being a constant. When $\gamma  = 0$ it reduces to the special case studied in
\cite{JCAP}. Since 
\bq
\lb{2.1h}
T_{\mu \nu}^{+}=0 ,  \;\;\; T_{\mu \nu}^{-}=\Lambda g^{-}_{\mu \nu},
\eq
we find
\bq
\lb{2.1i}
[T_{\mu \nu} u^{\mu}n^{\nu}]= T_{\mu \nu}^{+}u^{+ \mu}n^{+ \nu} - 
T_{\mu \nu}^{-}u^{- \mu}n^{- \nu} = 0.
\eq
Then, from (\ref{2.1d}) we find
\bq
\lb{2.1k}
\frac{\dot{\sigma}}{\sigma}=-2(2-\gamma) \frac{\dot{a}}{a},
\eq
which has the general solution,
\bq
\lb{2.1l}
\sigma=\sigma_{0}(\frac{a_{0}}{a})^{2(2-\gamma)}.
\eq

Setting
\bq
\lb{2.12}
{\cal{M}} = {m}{L_{0}}, \;\;\; 
a(\tau) = {R(\tau)}{L_{0}},\;\;\;
l = {L}{L_{0}}, 
\eq
where 
\bq
\lb{2.13}
L_{0} \equiv \left(4\pi \sigma_{0}a_{0}^{2(2-\gamma)}\right)^{-\frac{1}{2\gamma - 3}},
\eq
we find that Eq.(\ref{2.1a})  can be cast in the form,
\bq
\lb{2.1w}
 \frac{1}{2}{R^{*}}^{2}+V(R,m,L,\gamma)=0,
\eq
where $R^{*} \equiv dR/d(L^{-1}_{0}\tau)$,   and  
\bq
\lb{2.1z}
 V(R,m,L,\gamma)=-\frac{1}{2}\left\{-1+\frac{m}{R}
 +\frac{1}{4}{R^{4\gamma -6}} +m^{2}{R^{4(1-\gamma)}}
 +\frac{R^{2}}{2 L^{2}}-\frac{mR^{7-4 \gamma}}{L^{2}}+\frac{R^{10-4 \gamma}}{4L^{4}} \right\}.
\eq

Therefore, for any given constants $m$, $L$ and $\gamma$, Eq.(\ref{2.1w}) uniquely determines the collapse 
of the prototype  gravastar. Depending on the initial value $R_{0}$,  the collapse can
form either a black hole, or gravastar,  or a Minkowski, or a de Sitter space. 
In the last case, the thin shell
first collapses to a finite non-zero minimal radius and then expands to infinity.  To  guarantee  
that initially the spacetime does not have any kind of horizons,  cosmological or event,
we must restrict $R_{0}$ to the range,
\bq
\lb{2.2b}
2m < R_{0} < L,
\eq
correspondingly $a_{0} \in (2{\cal{M}}, l)$. When $m = 0= \Lambda$, the thin shell disappears,
and the whole spacetime is Minkowski. So, in the following we shall not consider this case.

From Eq.(\ref{2.1z}), we find that 
\bqn
\lb{nr1.1d}
\frac{\partial V(R,m,L,\gamma)}{\partial R} &=&  
\frac{m}{2}\left(\frac{1}{R^{2}}+ \left(7-4\gamma\right){\frac{{R}^{6-4\gamma}}{{L}^{2}}}\right) 
 - 2m^{2}\left(1-\gamma\right) {R}^{3-4\gamma} - {\frac {R}{2{L}^{2}}} \nonumber \\
& &  + \frac{1}{4}\left(3-2\gamma\right) {R}^{4\gamma-7}.
\eqn 
 
As both equations,  $V=0$  and ${\partial V(R,m,L,\gamma)}/{\partial R} = 0$, are quadratic in $m$, 
we can easily find $m$ from these two equations, which is given by  
\bqn
\lb{nr1.1da}
m_{c}(R,L,\gamma)&=&\frac{1}{2{{L}^{2} \left( -5\,{L}^{2}{R}^{4\,\gamma}+4\,{L}^{2}\gamma\,{R
}^{4\,\gamma}-3\,{R}^{8} \right) }}\,(-8\,{R}^{4\,\gamma+1}{L}^{4}+8\,{R}^{4\,\gamma+1}{L}^{4}
\gamma \nonumber \\ 
& & - 3\,{R}^{11}- 4\,{R}^{8\,\gamma-5}{L}^{4}\gamma+5\,{R}^{8\,\gamma -5}{L}^{4} \nonumber \\  
& & + 2\,{R}^{4\,\gamma+3}{L}^{2}-4\,{R}^{4\,\gamma+3}{L}^{2}\gamma),\;
(V = V' = 0).
\eqn
Substituting this expression  in $V = 0$ we find six functions $L(R,\gamma)$. 
Due to the complexity of these expressions we shall not give them here explicitly. Instead,
in the following we consider some particular cases.

%%%%%%%%%%%%%%%%%%%%%%%%%%%%%%%%%%%%%%%%%%%%%%%%%%%%%%%%%%%%%%%%%%%%%%%%%%%%%%%%%%%%%%%%%%%%%%%%%%%%
\subsection{$m = 0$}
%%%%%%%%%%%%%%%%%%%%%%%%%%%%%%%%%%%%%%%%%%%%%%%%%%%%%%%%%%%%%%%%%%%%%%%%%%%%%%%%%%%%%%%%%%%%%%%%%%%%

In this case, the spacetime outside the thin shell is flat, and the  mass of the shell completely
screens the  mass of the internal de Sitter spacetime. From Eq.(\ref{2.1z}) we find that 
\bq
\lb{nr1.6a}
V(R,L,\gamma)=-\frac{1}{2}\left\{-1 
 +\frac{1}{4}{R^{4\gamma -6}}  
 +\frac{R^{2}}{2 L^{2}} + \frac{R^{10-4 \gamma}}{4L^{4}} \right\}.
\eq
Then,   $V'(R) = 0$ yields,
\bq
\lb{vp}
\left(2\gamma - 3\right)L^{4} + 2 R^{4(2-\gamma)}L^{2} - \left(2\gamma - 5\right) R^{8(2-\gamma)} = 0,
\eq
which has real solution only for $\gamma < 3/2$ or $\gamma \ge 5/2$, and the corresponding solution
is given by
\bq
\lb{vp2}
L_{c} = \left|\frac{2\gamma - 5}{2\gamma - 3}\right|^{1/2} R_{c}^{2(2-\gamma)}, \; 
\gamma \not\in \left[{3}/{2}, {5}/{2}\right).
\eq
Substituting the above expression into the equation $V(R) = 0$, we find 
\bq
\lb{nr1.7a}
R_c  =  \left|\frac{2(\gamma - 2)}{2\gamma - 5}\right|^{\frac{1}{3 - 2\gamma}}.
\eq 

Figs. 1 and 2, and Table \ref{lcrc} show the functions of $R_{c}(\gamma)$ and
$L_{c}(\gamma)$, where for $\gamma \in [3/2, 5/2)$, the equations $V(R, L, \gamma)  = 0$
and $V'(R, L, \gamma)  = 0$ have no real solutions.  In particular,   when $\gamma=0.5$ we 
find that $L_{c}\approx 0.9185586537$ and 
$R_{c}\approx 0.8660254039$. For $L<L_{c}$ the potential is strictly negative as shown in 
Fig. 3. Thus, if the star starts to collapse at $R=R_{0}$, it will collapse 
continuously until $R=0$, whereby a Minkowski spacetime is formed. When $L=L_{c}$, since 
$R_{0}<L_{c}$, we can see that, the star will collapse until the center and turns the 
whole spacetime into a Minkowski. For $L>L_{c}$, the potential $V(R)$ is 
positive between $R_{1}$ and $R_{2}$, where $R_{1,2}$ are the two real roots of 
the equation $V(R,L>L_{c})=0$ with $R_{2}>R_{1}>0$. In this case, if the star starts 
to collapse with $R_{0}<R_{1}$, as can be seen from Fig. 3, it will 
collapse to $R=0$ whereby a Minkowski spacetime is finally formed. If it starts to
collapse with $R_{0}>R_{2}$, it will first collapse to $R=R_{2}$ and then starts
to expand until $R=\infty$, and the whole spacetime is finally de Sitter.

It should be noted that in this case we always have $V''(R_{c}, L_{c}, \gamma) < 0$,
that is, no stable stars exist.

\begin{table}
\caption{Some values of $R_{c}$ and $L_{c}$ as a function of $\gamma$  for
$\Lambda \not= 0 $ and $m = 0$.}
\label{lcrc}
    \centering
        \begin{tabular}{c| c c} \hline
     $\gamma$ & $R_{c}$ & $L_{c}$ \\ \hline
     $0.0$ & $0.9283177667$ & $0.9587624669$ \\
     $0.1$ & $0.9199519160$ & $0.9535623390$ \\
     $0.2$ & $0.9100298584$ & $0.9473109395$ \\
     $0.3$ & $0.8981404718$ & $0.9397043233$ \\
     $0.4$ & $0.8837277541$ & $0.9303202778$ \\
     $0.5$ & $0.8660254039$ & $0.9185586537$ \\
     $0.6$ & $0.8439546889$ & $0.9035435931$ \\
     $0.7$ & $0.8159607706$ & $0.8839575014$ \\
     $0.8$ & $0.7797433156$ & $0.8577474304$ \\
     $0.9$ & $0.7318025503$ & $0.8215821352$ \\
     $1.0$ & $0.6666666667$ & $0.7698003592$ \\
     $1.1$ & $0.5756295655$ & $0.6922956261$ \\
     $1.2$ & $0.4452231361$ & $0.5703420705$ \\
     $1.3$ & $0.2598914458$ & $0.3713508347$ \\
     $1.4$ & $0.0482828420$ & $0.08734694771$ \\
     $2.6$ & $0.4428889922$ & $0.8012163331$ \\
     $2.7$ & $0.5933418502$ & $0.8478077860$ \\
     $2.8$ & $0.6857500883$ & $0.8784631649$ \\
     $2.9$ & $0.7485495080$ & $0.9002622198$ \\
     $3.0$ & $0.7937005260$ & $0.9164864251$ \\
     $5.0$ & $0.9742903329$ & $0.9881107444$ \\
    \hline
       \end{tabular}
\end{table}
 
\begin{figure}
\label{Rcm0a}
\centering
\includegraphics[width=12cm]{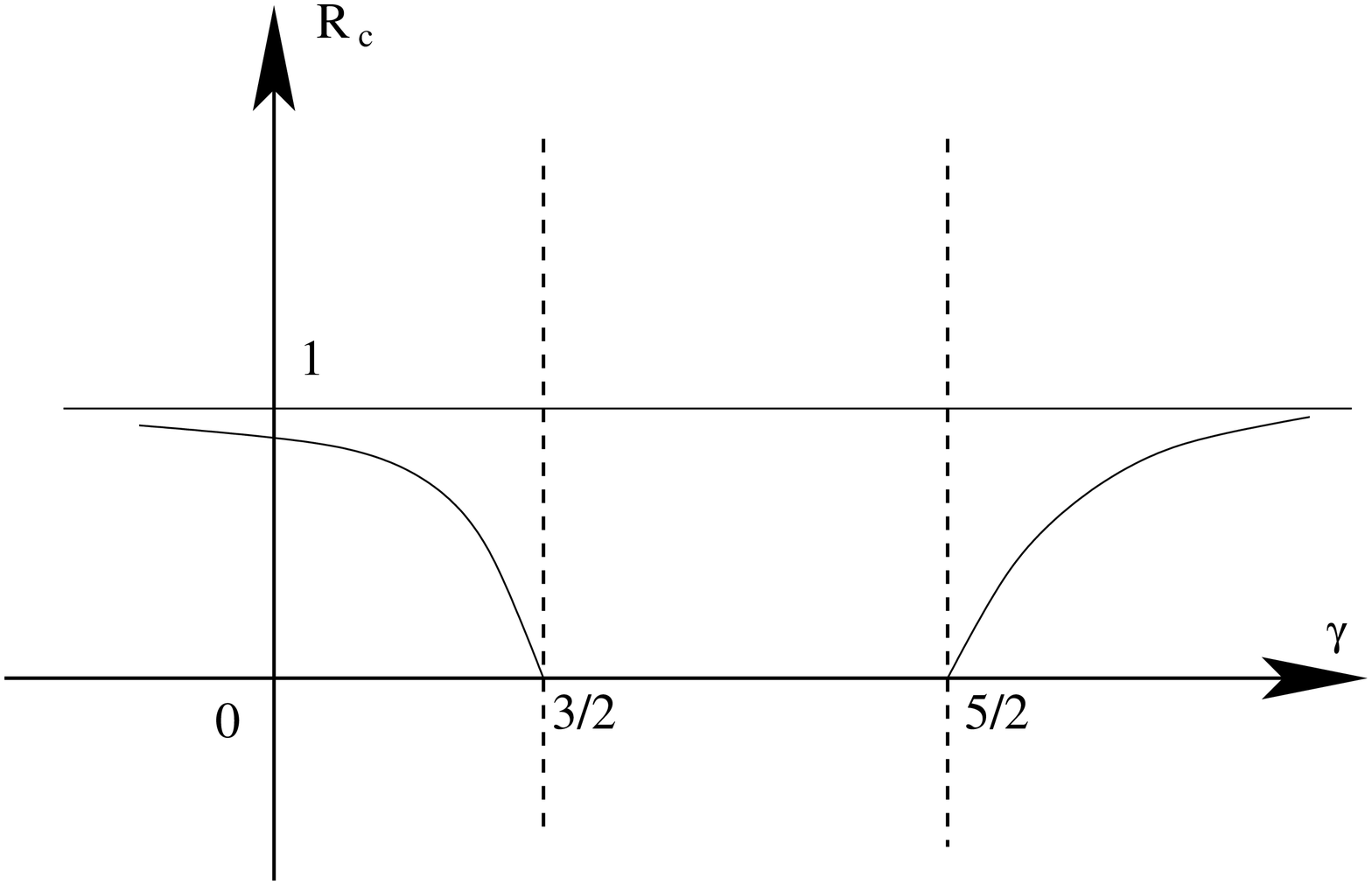}
\caption{The function $R_{c}$ defined in Eq.(\ref{vp}) for $m=0$.} 
\end{figure}

%Both Table \ref{lcrc} and Fig. 1 and 2 show the dependence of $R_{c}$ and $L_c$ on $\gamma$. 

\begin{figure}
\label{Lcm0a}
\centering
\includegraphics[width=12cm]{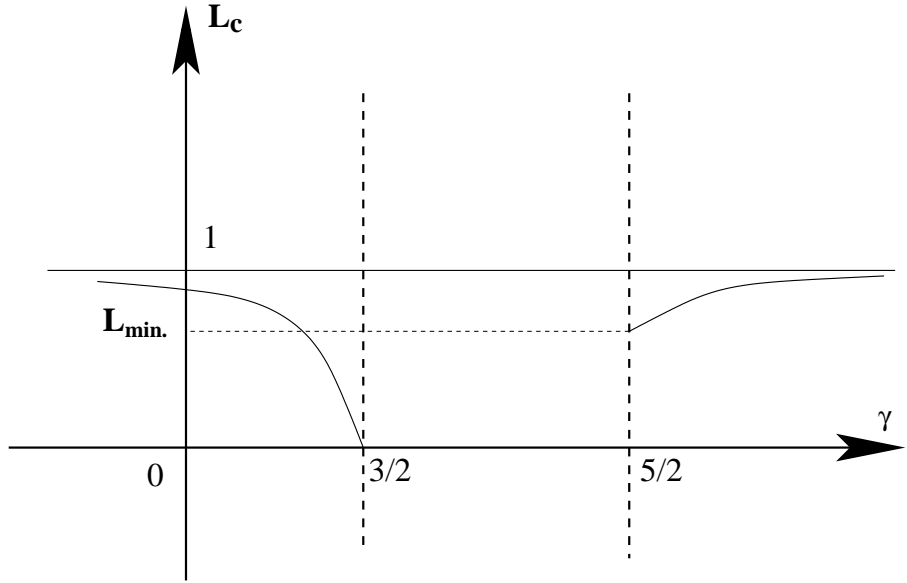}
\caption{The function $L_{c}$ defined in Eq.(\ref{vp2}) for $m=0$, where
$L_{min} = 1/\sqrt{2}$.} 
\end{figure}

\begin{figure}
\label{gamma0v5}
\centering
\includegraphics[width=12cm]{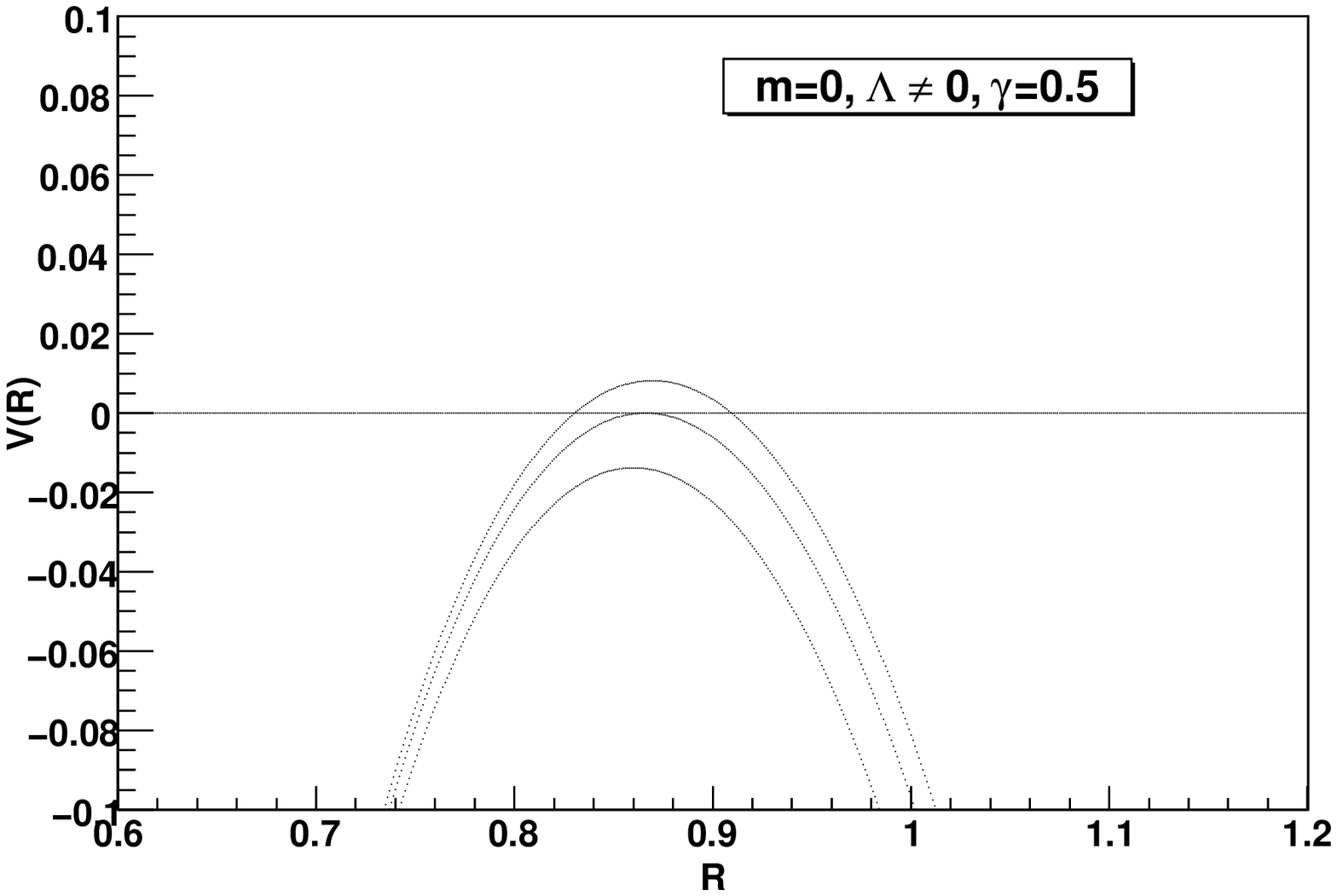}
\caption{The potential V(R) for $m=0$ and $\gamma=0.5$. The top line is for $L>L_{c}$, 
the middle line  is for $L=L_{c}$, and the bottom line  is for $L<L_{c}$.}
\end{figure}

%Lc is imaginary for m=0 and \gamma>3/2
%\begin{figure}
%\label{gamma3v0}
%\centering
%\includegraphics[width=12cm]{gamma3v0.eps}
%\caption{The potential V(R) for $\gamma=3.0$. The top line is for $L>L_{c}$, the middle line is
%$L=L_{c}$, and the bottom line is for $L<L_{c}$.}
%\end{figure}

%%%%%%%%%%%%%%%%%%%%%%%%%%%%%%%%%%%%%%%%%%%%%%%%%%%%%%%%%%%%%%%%%%%%%%%%%%%%%%%%%%%%%%%%%%%%%%%%%%%%
\subsection{$\Lambda = 0$}
%%%%%%%%%%%%%%%%%%%%%%%%%%%%%%%%%%%%%%%%%%%%%%%%%%%%%%%%%%%%%%%%%%%%%%%%%%%%%%%%%%%%%%%%%%%%%%%%%%%%

In this case, Eq.(\ref{2.1z}) reduces to
\bq
\lb{nr1.3}
V(R,m,\gamma)= \frac{1}{8{R}^{6}} \left(4\,{R}^{6}-4\,m{R}^{5}-{R}^{4\,\gamma}-4\,{R}^{10-4
\gamma}{m}^{2}\right),
\eq
from which we find that the equations $V(R)=0$ and $V'(R)=0$ have the explicit solutions,
\bqn
\lb{nr1.4c}
R_{c}(\gamma)&=&\left|{\frac{4-4\,\gamma}{5-4\,\gamma}}\right|^{\frac{1}{\left( 2\,\gamma-3 \right)}},\\
\lb{nr1.4j}
m_{c}&=&\frac{1}{2} \sqrt{ R_c^{2(4\gamma-5)} + \frac{4}{5-4\gamma} R_c ^ {4\gamma -4}}.
\eqn

\begin{figure}
\label{RcLinf}
\centering
\includegraphics[width=12cm]{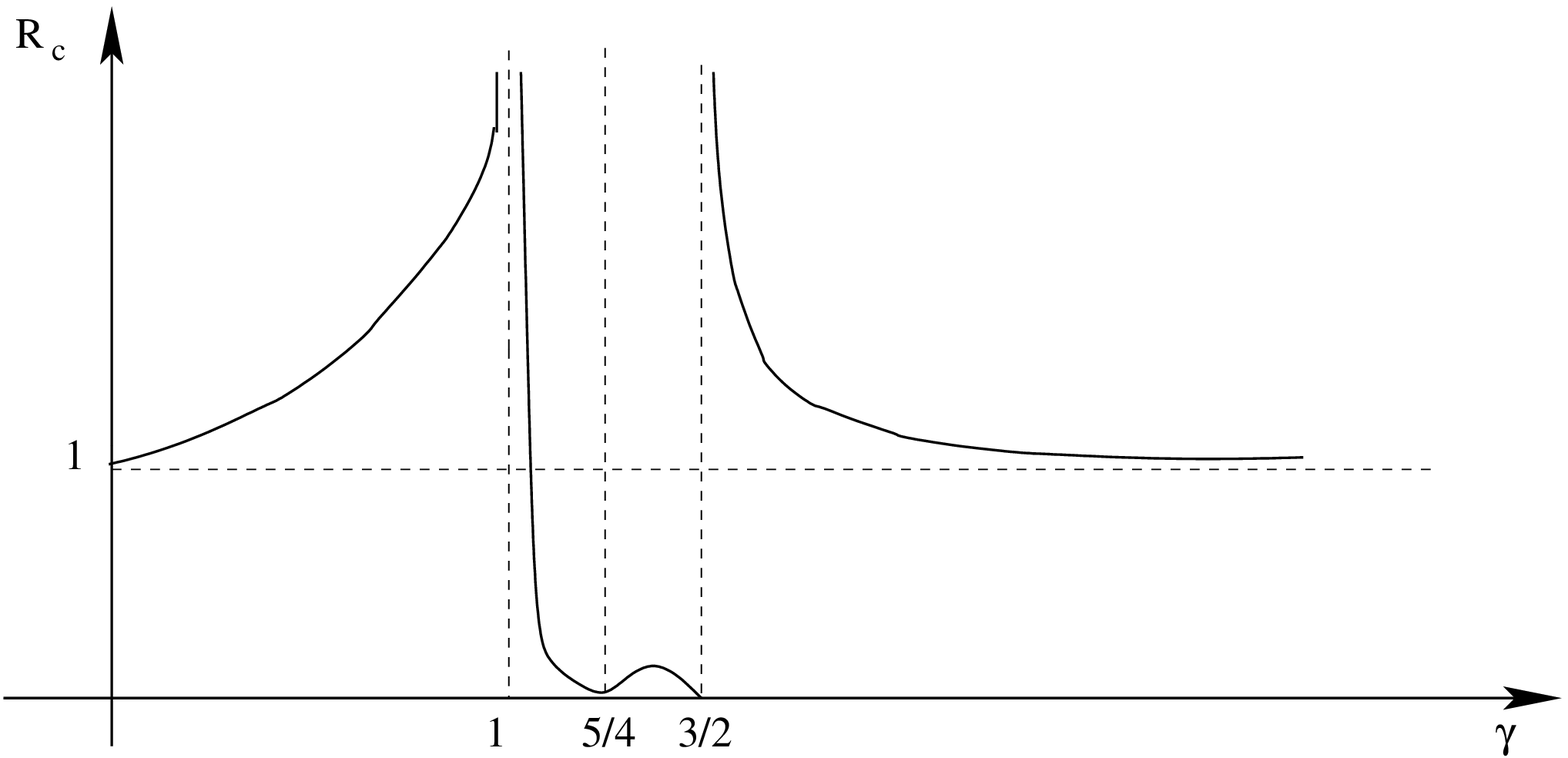}
\caption{The function $R_{c}$ defined in Eq.(\ref{nr1.4c}) for $\Lambda=0$.}
\end{figure}

\begin{figure}
\label{mcLinf}
\centering
\includegraphics[width=12cm]{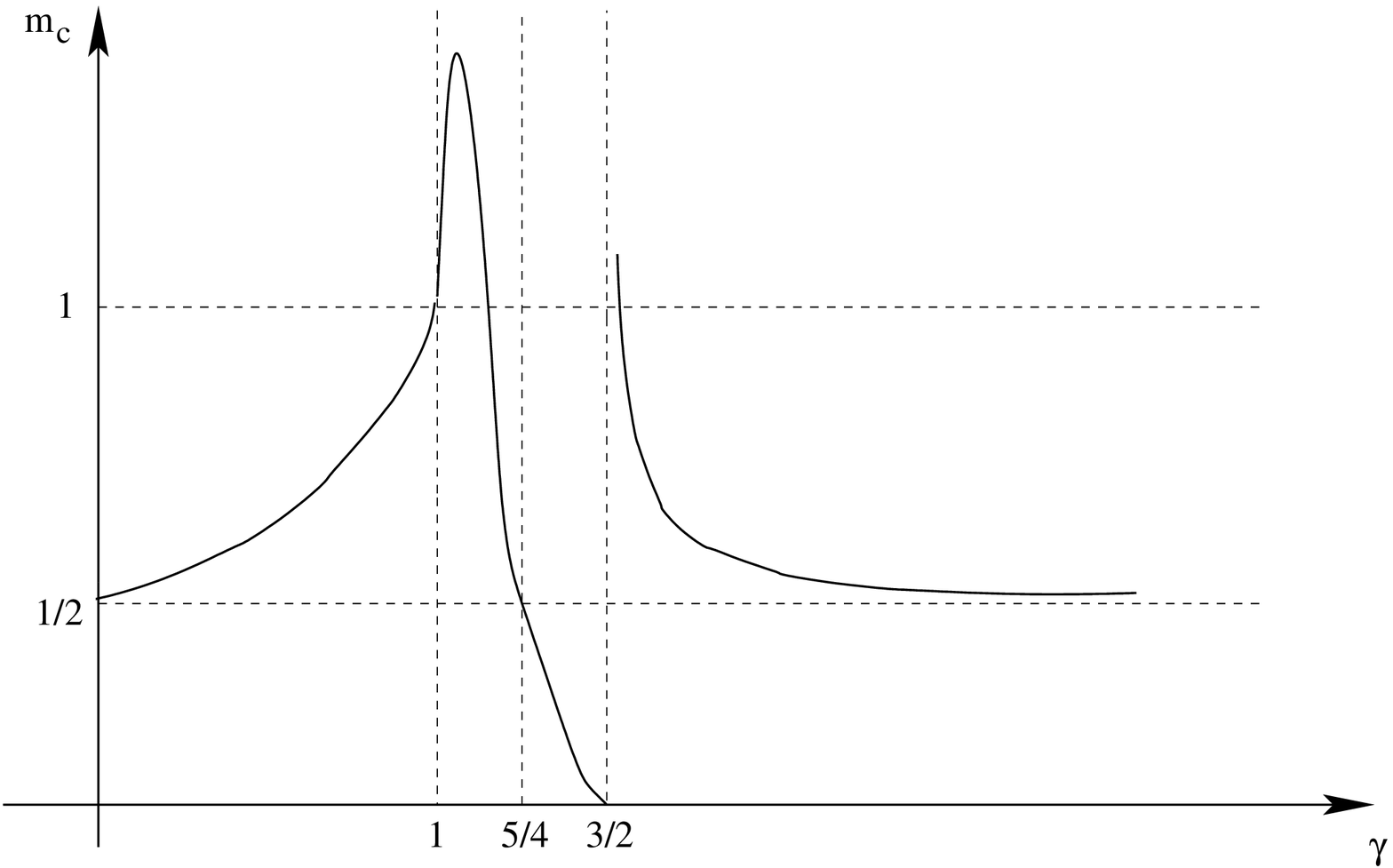}
\caption{The function $m_{c}$ defined in Eq.(\ref{nr1.4j}) for $\Lambda=0$.   }
\end{figure}

\begin{figure}
\label{gammamnotzero0v5}
\centering
\includegraphics[width=12cm]{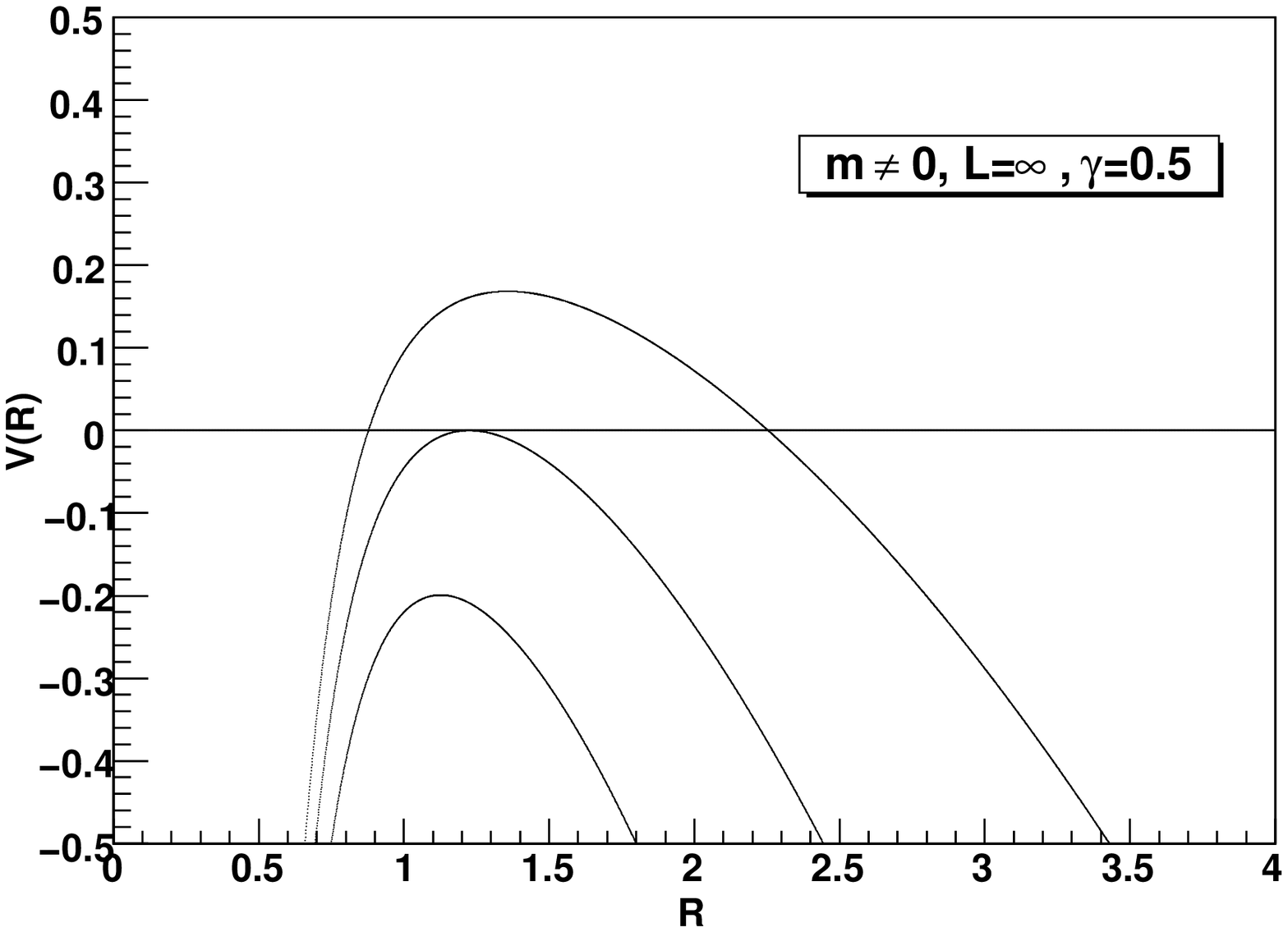}
\caption{The potential V(R) for $\gamma=0.5$. The top line is for $m<m_{c}$, 
the middle line is for $m=m_{c}$, and the bottom line is for $m>m_{c}$.}
\end{figure}

\begin{figure}
\label{gammamnotzero3v0}
\centering
\includegraphics[width=12cm]{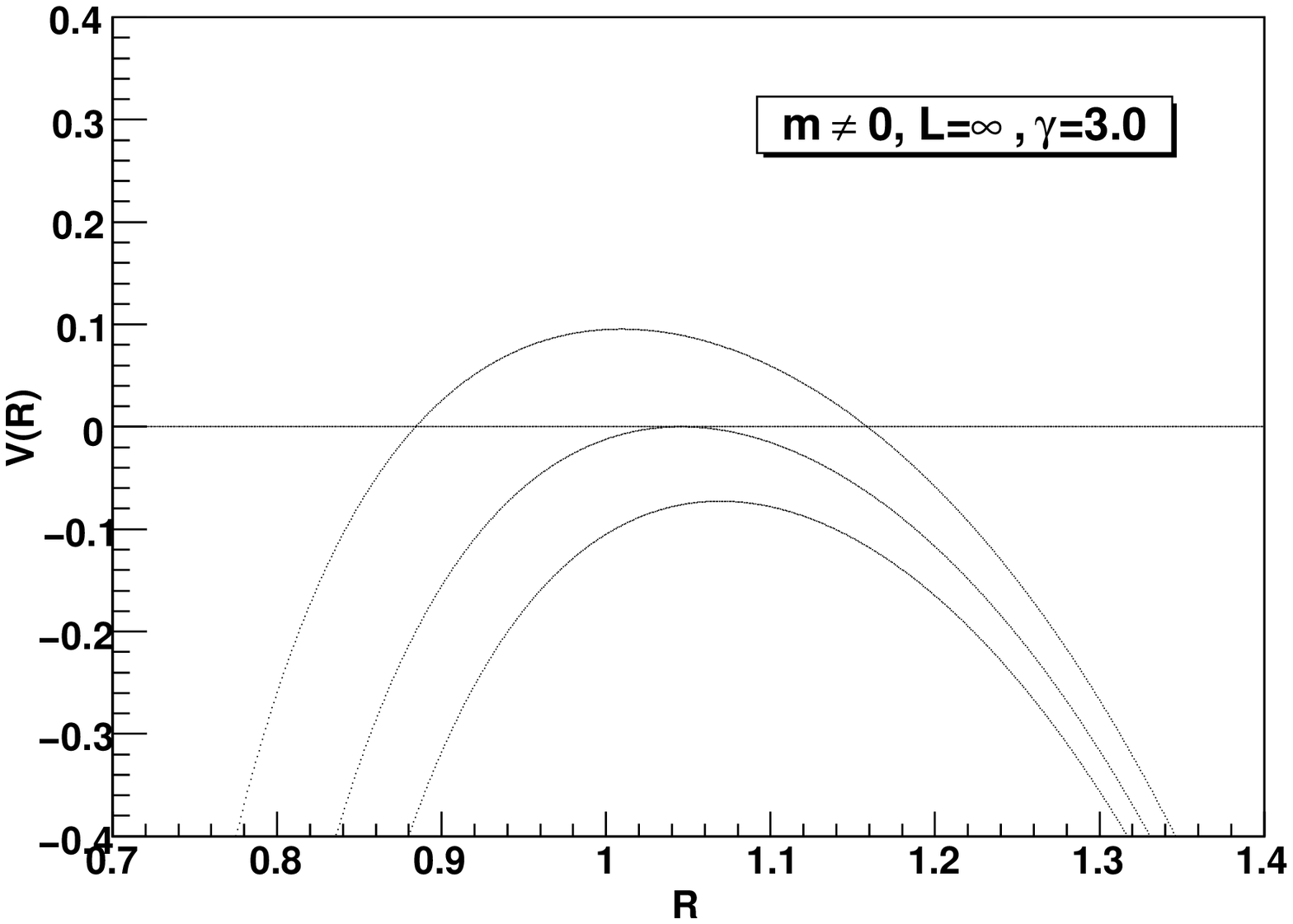}
\caption{The potential V(R) for $\gamma=3.0$. The top line is for $m<m_{c}$, 
the middle line is for $m=m_{c}$, and the bottom line is for $m>m_{c}$.}
\end{figure}

\begin{table}
\label{rcmc}
\caption{Some values of $R_{c}$ and $m_{c}$ as a function of $\gamma$ for $\Lambda = 0 $ and $m \not= 0$}
	\centering
		\begin{tabular}{c| c c} \hline
     $\gamma$ & $R_{c}$ & $m_{c}$ \\ \hline
     $0.0$ & $1.077217345$ & $0.5170643255$ \\ 
     $0.1$ & $1.091490000$ & $0.5199536485$ \\ 
     $0.2$ & $1.110255191$ & $0.5236577775$ \\ 
     $0.3$ & $1.135692224$ & $0.5285215890$ \\ 
     $0.4$ & $1.171542463$ & $0.5350989795$ \\ 
     $0.5$ & $1.224744871$ & $0.5443310540$ \\ 
     $0.6$ & $1.309606309$ & $0.5579387820$ \\ 
     $0.7$ & $1.460581840$ & $0.5794043665$ \\ 
     $0.8$ & $1.784674184$ & $0.6169244095$ \\ 
     $0.9$ & $2.840468889$ & $0.6956250340$ \\
     $1.0$ & $\infty$      & 1.000000000  \\
     $1.1$ & $1.475575893$ & $1.660022879$ \\
     $1.2$ & $0.09921256570$ & $1.190550789$ \\
     $1.3$ & $0.01134023029$ & $0.1360827636$ \\
     $1.4$ & $0.007415771480$ & $0.006591796870$ \\
     $1.5$ & Undefined     & Undefined  \\
     $1.6$ & $3.625777848$ & $14.80525970$ \\
     $1.7$ & $3.017962339$ & $1.043246245$ \\
     $1.8$ & $1.867291683$ & $0.7407438105$ \\  
     $1.9$ & $1.501970633$ & $0.6398928165$ \\  
     $2.0$ & $1.333333333$ & $0.5925925920$ \\  
     $3.0$ & $1.045515917$ & $0.5120894280$ \\  
     $4.0$ & $1.017554577$ & $0.5045725165$ \\  
     $1.0$x$10^{5}$& $1.000000000$ & $0.5000000000$ \\  
     \hline
   	\end{tabular}
\end{table}

Figs. 4 and 5 show the dependence of $R_{c}$ and $m_{c}$ on $\gamma$. 
 The results of our previous work are obtained when $\gamma=0$ \cite{JCAP}. 
Some representative cases are $\gamma=0.5$ and $\gamma=3.0$.
For $\gamma=0.5$, we find that $m_{c}\approx 0.5443310540$ and $R_{c}\approx 1.224744871$,
and for $\gamma=3.0$ we find that $m_{c}\approx 0.5120894280$ and $R_{c}\approx 1.045515917$. 
In both cases, for $m>m_{c}$ the potential $V(R)$ is strictly negative as shown in Figs. 6 and 7. 
Then, the collapse always forms black holes. 
For $m=m_{c}$, there are two different possibilities, depending on the choice of the initial 
radius $R_{0}$. In particular, if the star begins to collapse with $R_{0}>R_{c}$, the collapse 
will asymptotically approach the minimal radius $R_{c}$. Once it collapses to this point, the 
shell will stop collapsing and remains there for ever.
However, in this case this point is unstable and any small perturbations will lead the star 
either to expand for ever and leave behind a flat spacetime, or to collapse until $R=0$, 
whereby a Schwarzschild black hole is finally formed. On the other hand, if the star begins 
to collapse with $2m_{c}<R_{0}<R_{c}$ as shown in Figs. 6 and 7, the star will collapse until a 
black hole is formed. For $m<m_{c}$, 
the potentials $V(R)$ for each case have a positive maximal, and the equation $V(R,m<m_{c})=0$ 
has two positive roots $R_{1,2}$ with $R_{2}>R_{1}>0$. There are two possibilities here, 
depending on the choice of the initial radius $R_{0}$. If $R_{0}>R_{2}$, the star will first
collapse to its minimal radius $R=R_{2}$ and then expand to infinity, whereby a Minkowski 
spacetime is finally formed. If $2m<R_{0}<R_{1}$, the star will collapse continuously until
$R=0$, and a black hole will be finally formed.
 
 It should be noted that, similar to the last case, now we always have $V''(R_{c}, m_{c}, \gamma) < 0$,
which means that  no stable stars exist in this case, too.

%%%%%%%%%%%%%%%%%%%%%%%%%%%%%%%%%%%%%%%%%%%%%%%%%%%%%%%%%%%%%%%%%%%%%%%%%%%%%%%%%%%%%%%%%%%%%%%%%
\subsection{$m \not= 0 $ and $\Lambda \not= 0$}
%%%%%%%%%%%%%%%%%%%%%%%%%%%%%%%%%%%%%%%%%%%%%%%%%%%%%%%%%%%%%%%%%%%%%%%%%%%%%%%%%%%%%%%%%%%%%%%%%

\begin{table}
\label{generalcase}
\caption{Some values of $m_{c}$ and $L_{c}$ obtained numerically as a function of $\gamma$ in 
the general case where $\Lambda \not= 0$ and $m \not= 0$}
	\centering
\begin{tabular}{c| c c} \hline
     $\gamma$ & $m_{c}$ & $L_{c}$ \\ \hline
     $0.0$ & $0.5170643255$ & $2.8743398$\\
     $0.1$ & $0.5199536485$ & $3.1000618$\\ 
     $0.2$ & $0.5236577775$ & $3.3917341$ \\ 
     $0.3$ & $0.5285215890$ & $3.7828681$ \\
     $0.4$ & $0.5350989795$ & $4.3336020$ \\
     $0.5$ & $0.5443310540$ & $5.1626297$ \\
     $0.6$ & $0.5579387820$ & $6.5372013$ \\
     $0.7$ & $0.5794043665$ & $9.1891232$ \\
     $0.8$ & $0.6169244095$ & $15.8955019$ \\
     $0.9$ & $0.6956250340$ & $47.7590095$ \\
     $0.95$ & $0.7788797565$ & $166.7543222$ \\
     $0.99$ & $0.9200729380$ & $3906.8991705$ \\
     $0.991$ & $0.9259539055$ & $4823.955200$ \\
     $0.992$ & $0.9320895460$ & $6107.378535$ \\
     $0.993$ & $0.9385115510$ & $7981.5666867$ \\
     $0.994$ & $0.9452603485$ & $10873.01788$ \\
     $0.995$ & $0.9523894295$ & $15675.49526$ \\
     $0.996$ & $0.9599731720$ & $24531.28407$ \\     
     $0.997$ & $0.9681226815$ & $43701.09808$ \\          
     $0.998$ & $0.9770238615$ & $98596.55052$ \\          
     $0.999$ & $0.9870615640$ & $395877.877549$ \\          
     $0.9999$ & $0.9982370395$ & $0.3980357873\times 10^{8}$\\          
     $1.7$ & $1.043246242$ & $0.6064576241\times 10^{8}$\\               
     $3.0$ & $0.5120894280$ & $10410.51705$\\               
     $5.0$ & $0.5023884065$ & $6235.986909$\\                    
     \hline
   	\end{tabular}
\end{table}

As mentioned before, the analytic expression  for $L_{c}$ in the present case
are too complicated to write out here.
Instead, we shall study it numerically. Our main strategy is to start with 
$m_{c}$ obtained for the case $\Lambda=0$, and then gradually turn on
$\Lambda$.  We plot the potential $V(R,m_{c}(\gamma),L,\gamma)$ as a function of $R$ for any given
$\gamma$,  by finely tuning $L$ until a stable gravastar or a bounded excursion gravastar
 is found [see Figs. 8-11]. 
The value $L_{c}$, as shown in Table III, is  obtained numerically when 
a stable gravastar is found for a specific pair $(m_{c},\gamma)$. 

It can be shown that  both types of  gravastars  
can be formed for $\gamma \in [0, 1)$. But for $\gamma \ge 1$, we find that for any given values of $L$ and $m$ 
only black holes can be formed , for example, see  Figs. 19-24.
In particular, as $\gamma \rightarrow 1$, we find that $L_{c}\rightarrow\infty$.
This can be seen from Table III and Figs. 12-18.

\begin{figure}
\label{gravgamma0deltal}
\centering
\includegraphics[width=12cm]{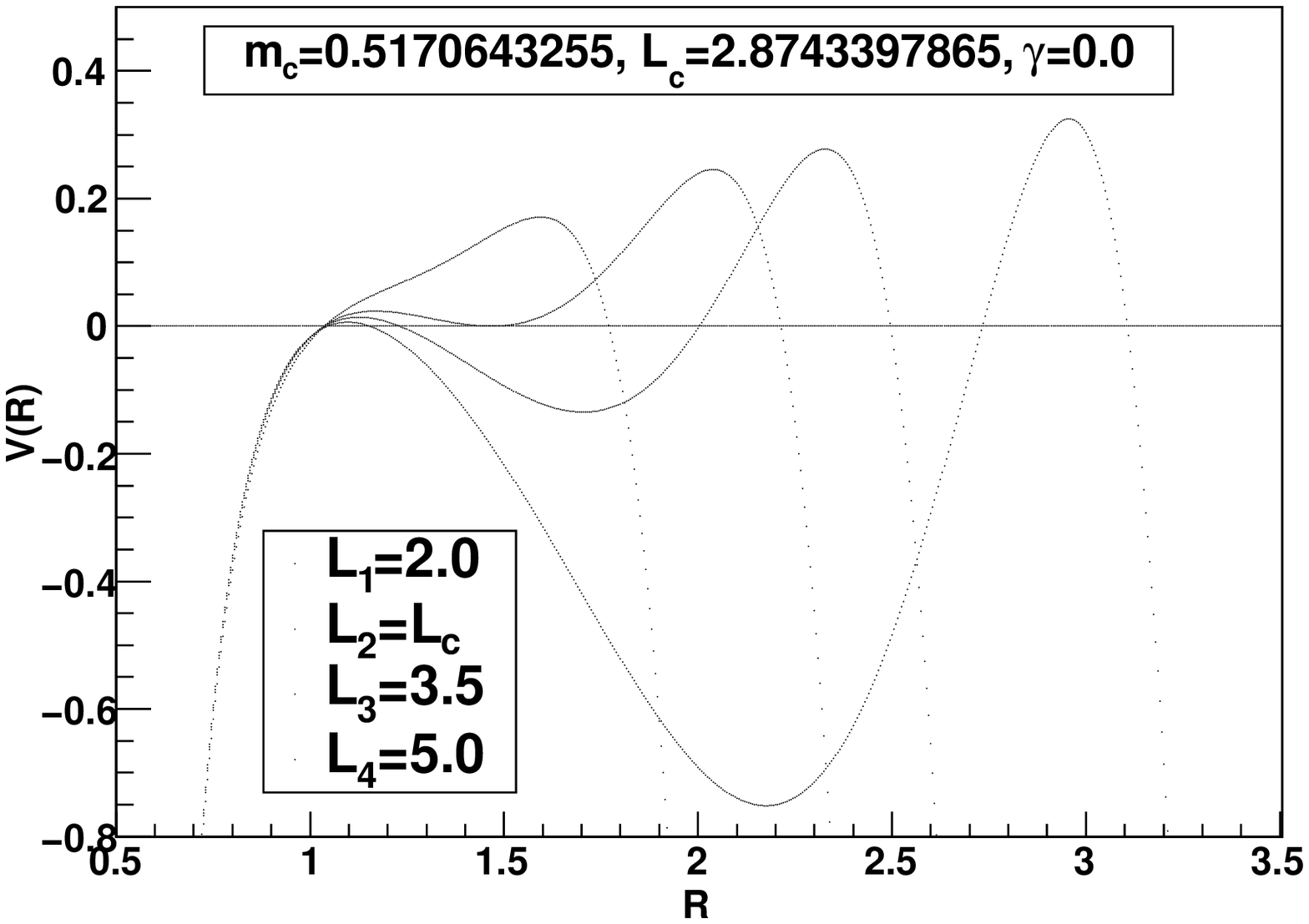}
\caption{The potential V(R) for $\gamma=0$ with some values of $ L$ near the
critical point  $L = L_{c}$. The curves from top to botton represent $L_1$ to $L_4$, respectively.}
\end{figure}

\begin{figure}
\label{gravgamma0deltam}
\centering
\includegraphics[width=12cm]{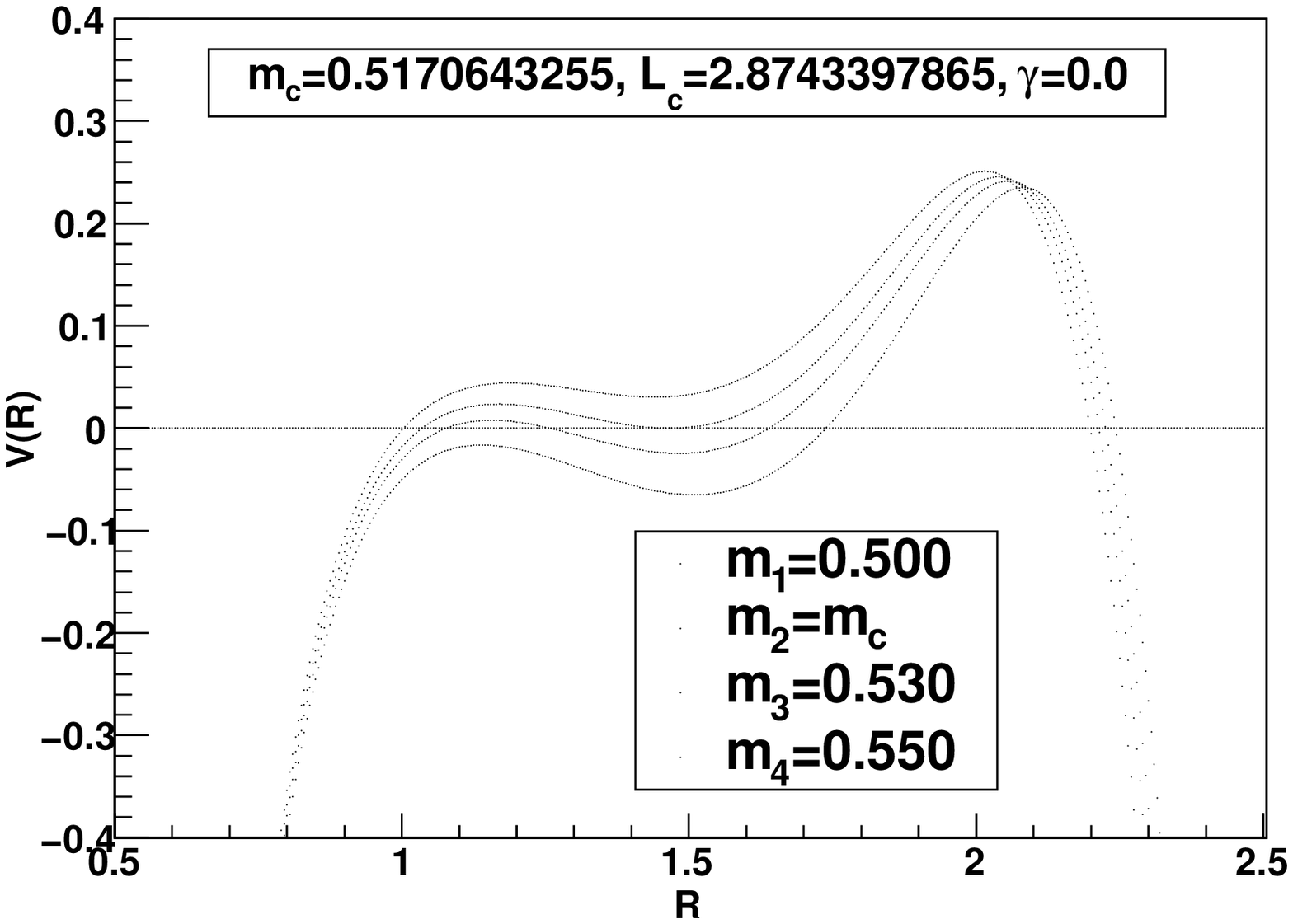}
\caption{The potential V(R) for $\gamma=0$ with some values of  $m$ near the
critical point $m = m_{c}$. The curves from top to botton represent $m_1$ to $m_4$, respectively.}
\end{figure}
 
\begin{figure}
\label{gravgamma0v4deltal}
\centering
\includegraphics[width=12cm]{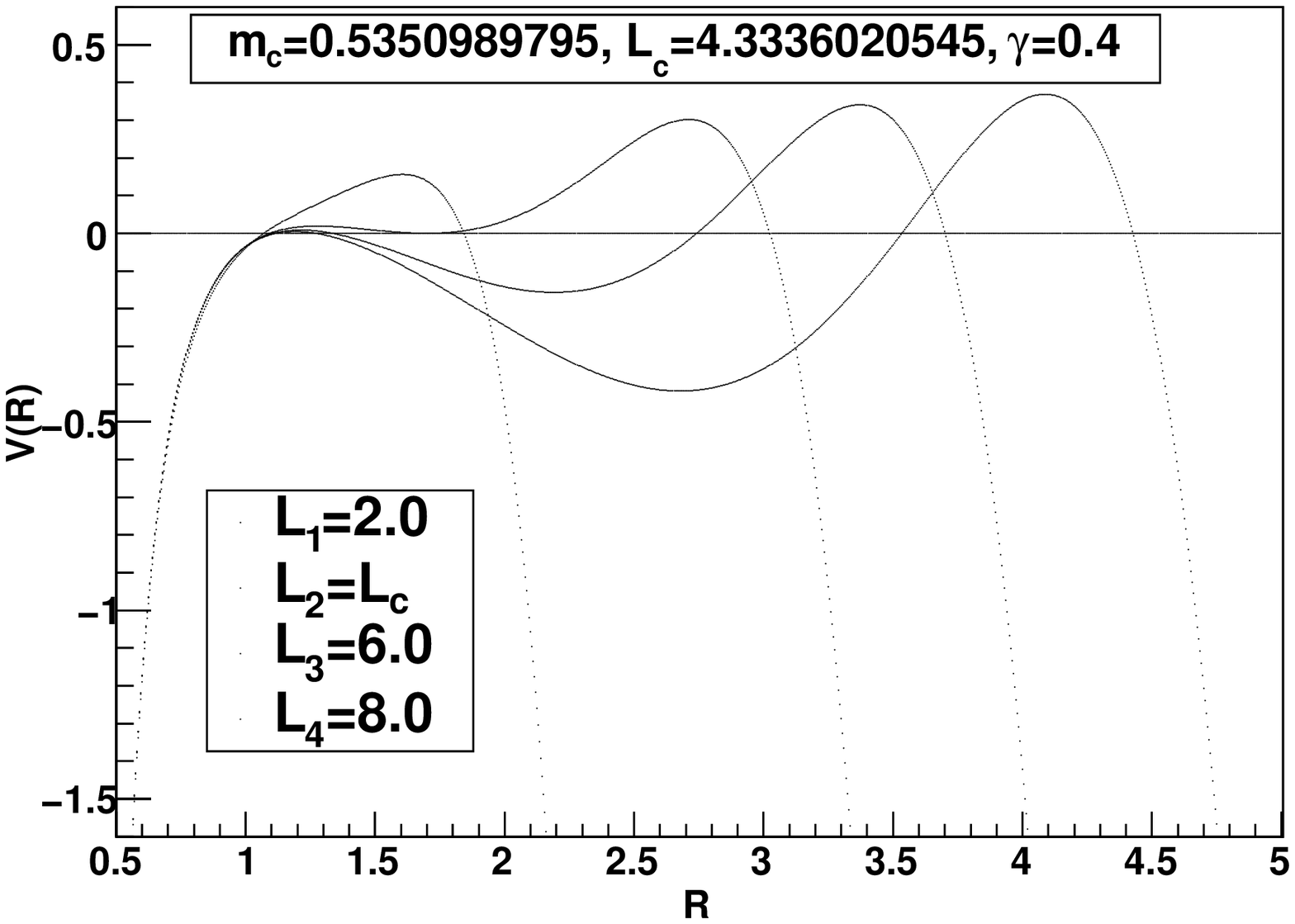}
\caption{The potential V(R) for $\gamma=0.4$ with  some values of  $ L$ near the
critical point  $L = L_{c}$. The curves from top to botton represent $L_1$ to $L_4$, respectively.}
\end{figure}

\begin{figure}
\label{gravgamma0v4deltam}
\centering
\includegraphics[width=12cm]{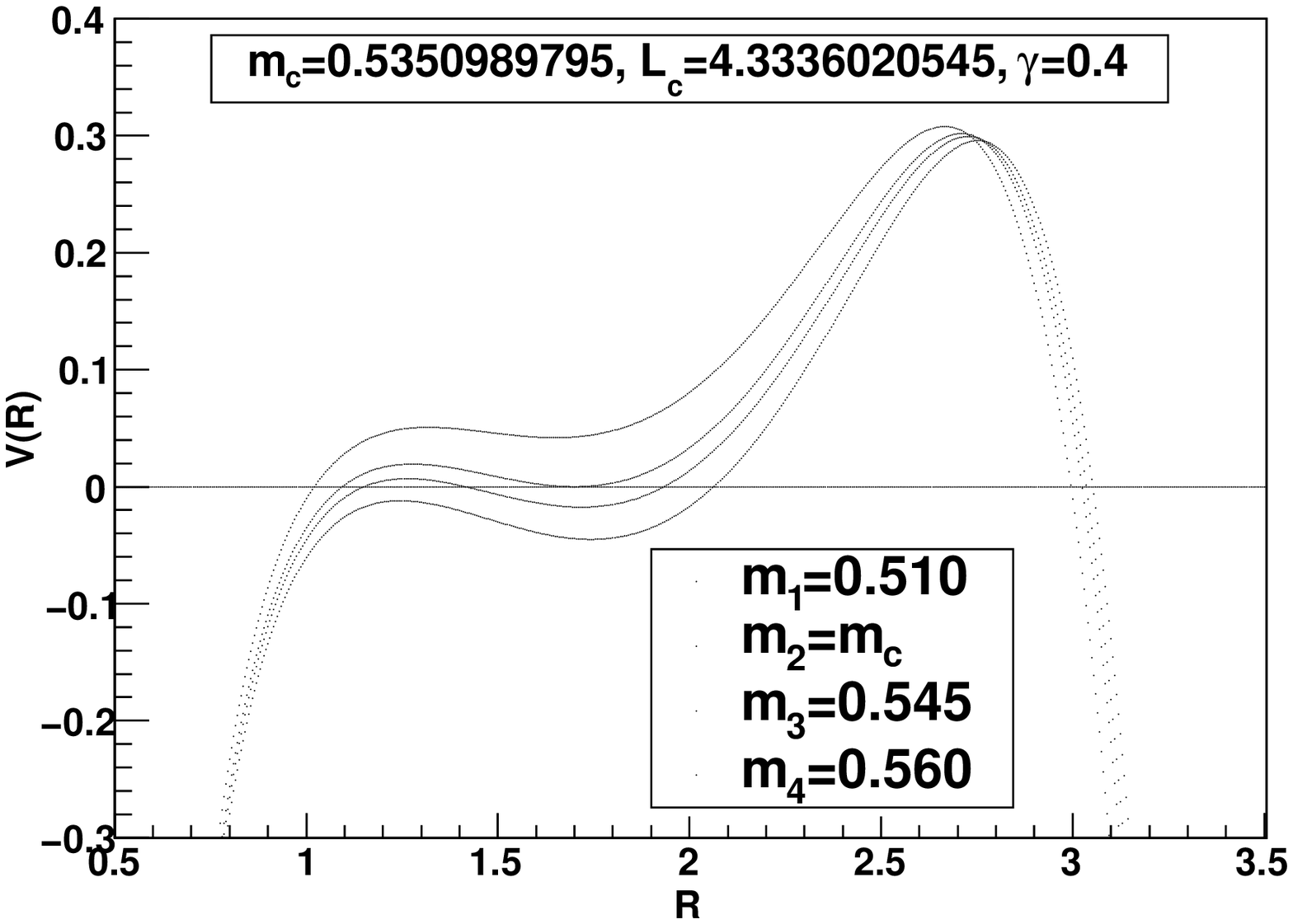}
\caption{The potential V(R) for $\gamma=0.4$ with some values of  $ m$ near the
critical point $ m = m_{c}$. The curves from top to botton represent $m_1$ to $m_4$, respectively.}
\end{figure}

\begin{figure}
\label{generalcasegamma0v7}
\centering
\includegraphics[width=12cm]{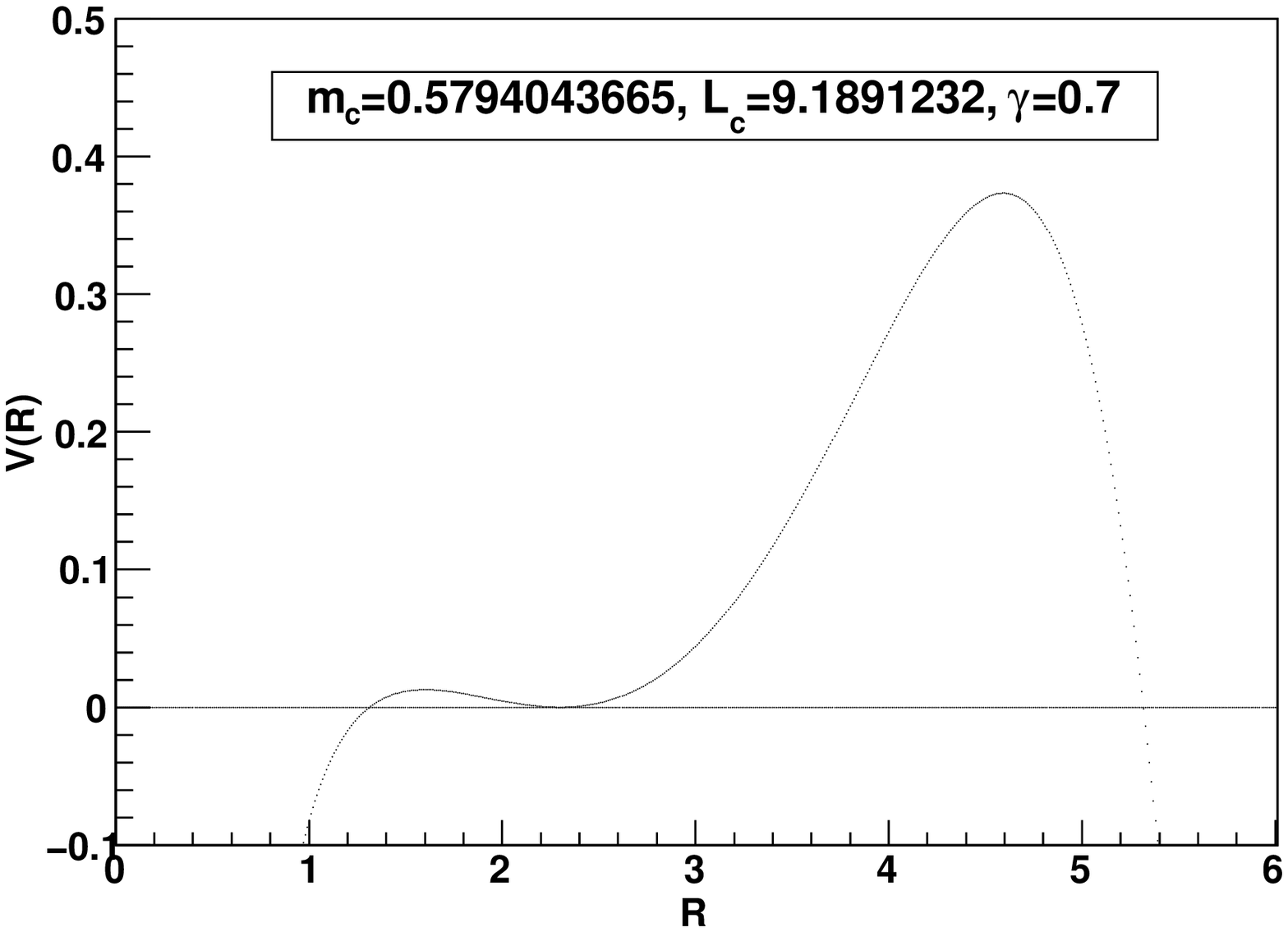}
\caption{The potential V(R) for $\gamma=0.7$}
\end{figure}

%\newpage

\begin{figure}
\label{generalcasegamma0v8}
\centering
\includegraphics[width=12cm]{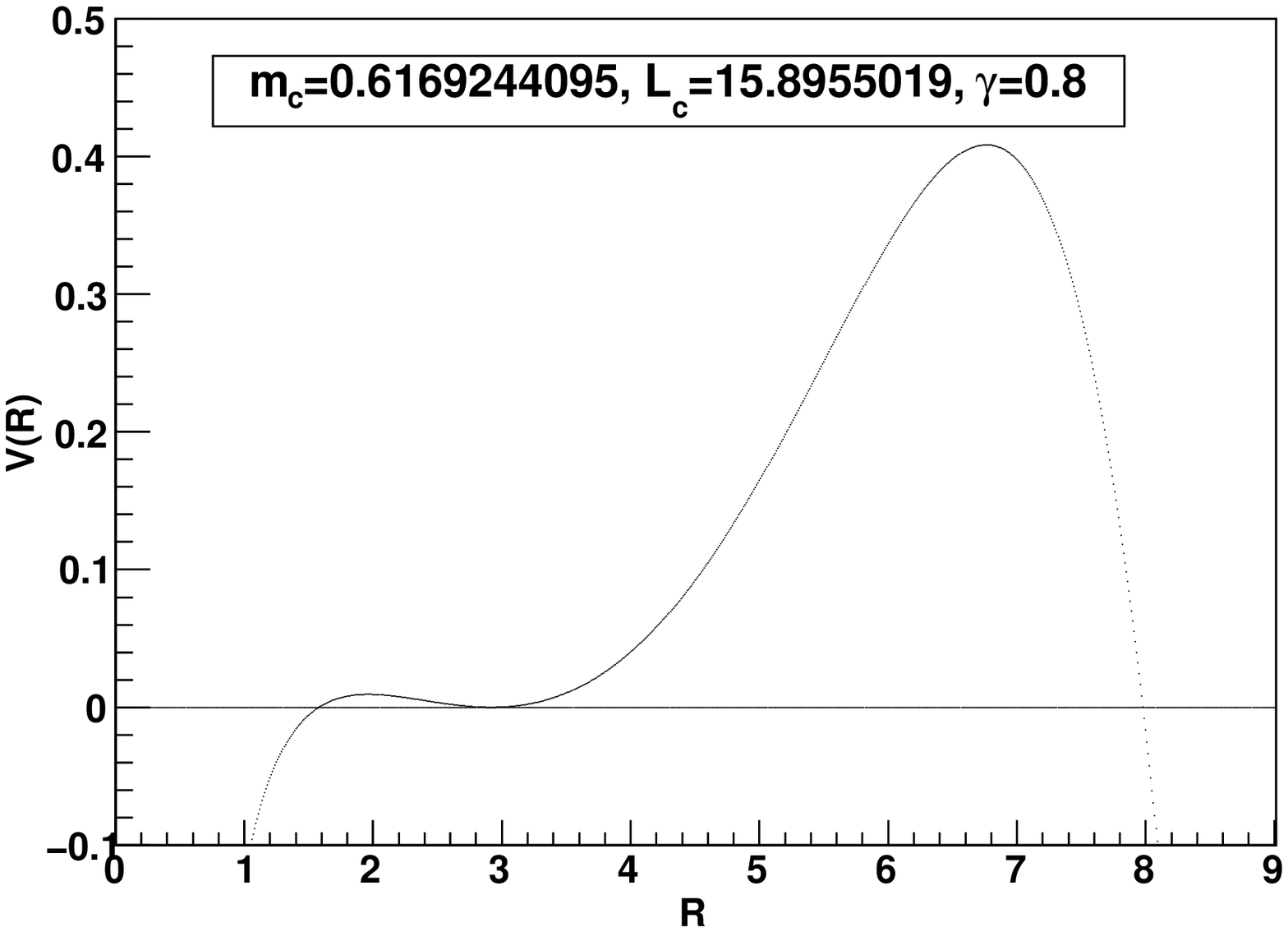}
\caption{The potential V(R) for $\gamma=0.8$.}
\end{figure}

\begin{figure}
\label{generalcasegamma0v9}
\centering
\includegraphics[width=12cm]{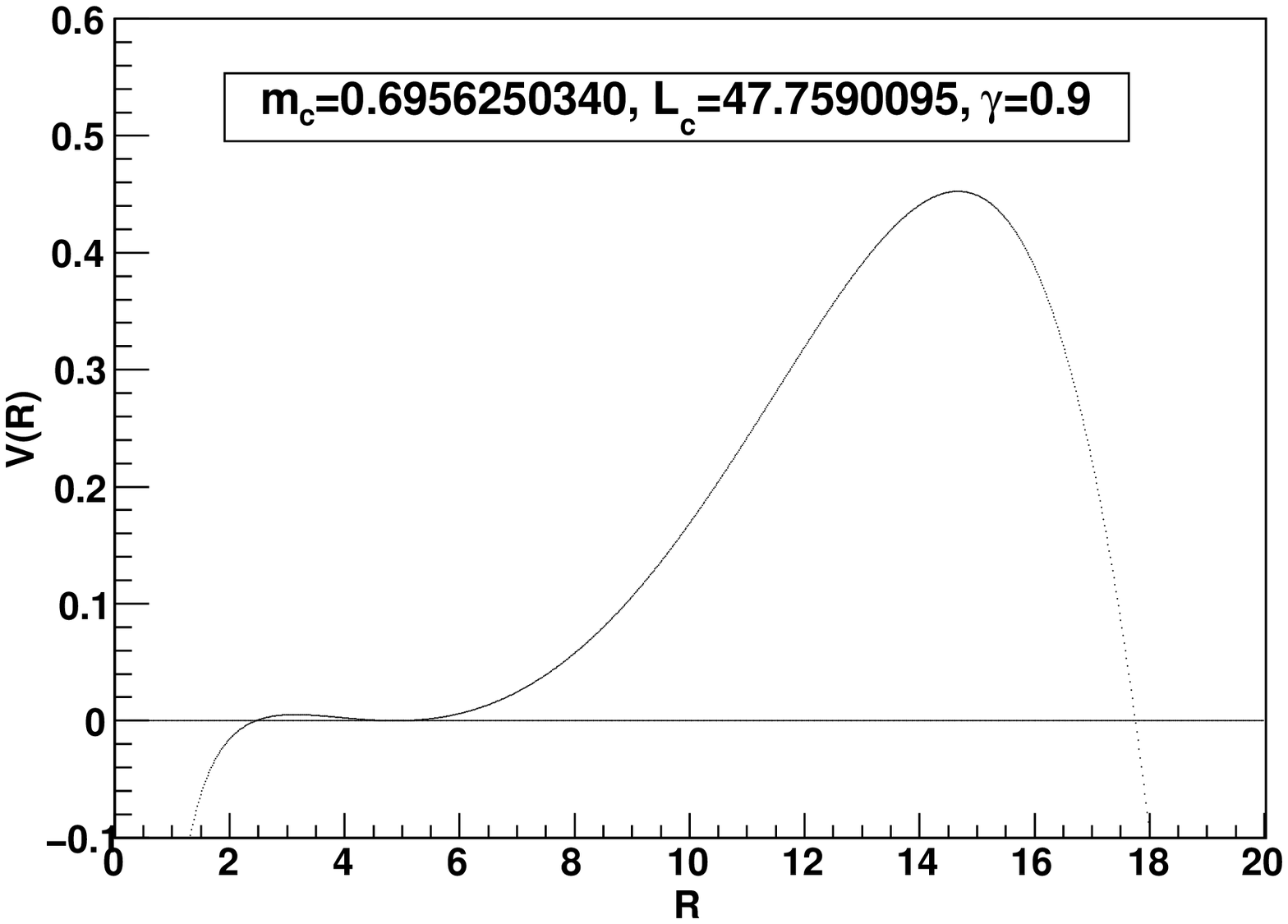}
\caption{The potential V(R) for $\gamma=0.9$.}
\end{figure}

\begin{figure}
\label{generalcasegamma0v95}
\centering
\includegraphics[width=12cm]{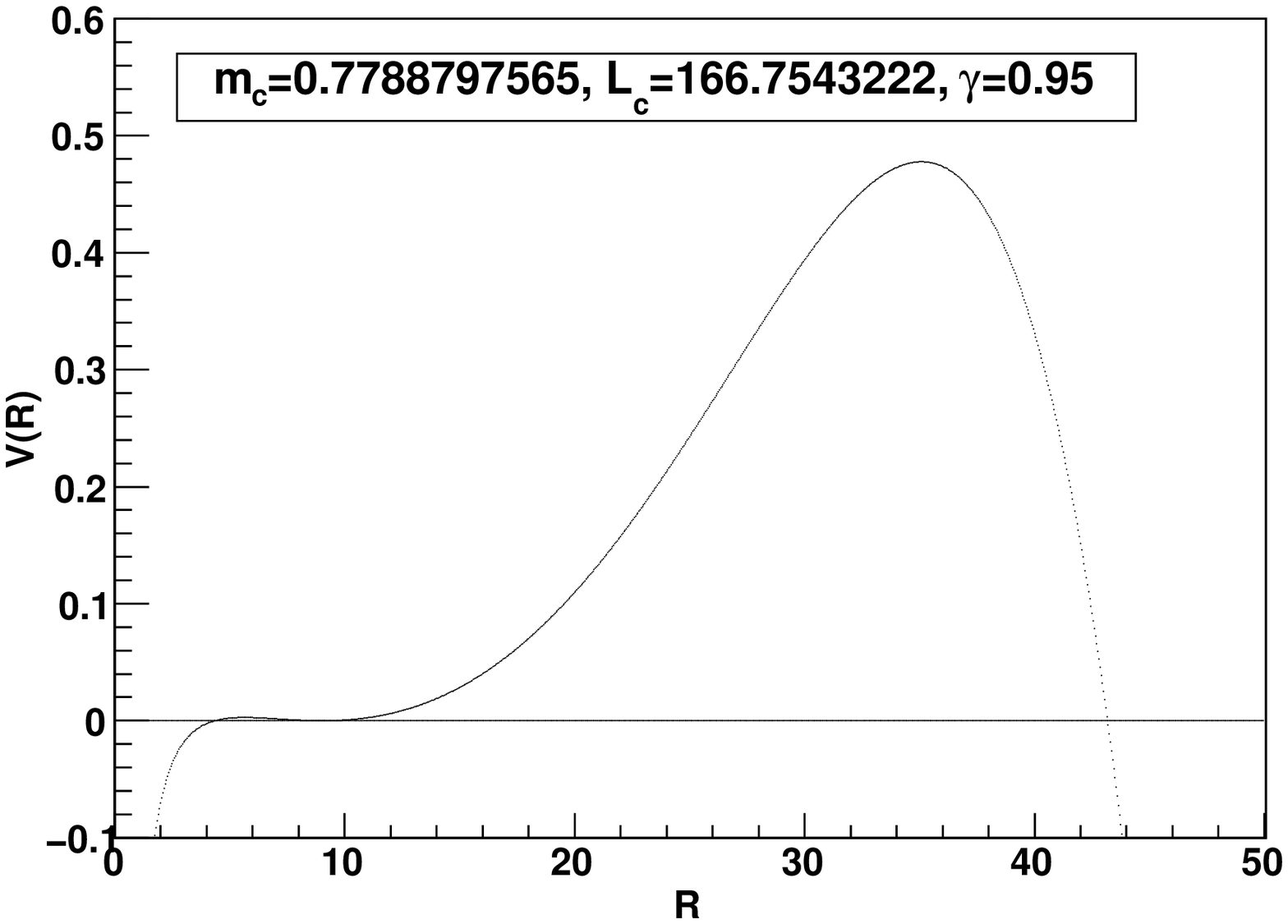}
\caption{The potential V(R) for $\gamma=0.95$.}
\end{figure}

\begin{figure}
\label{generalcasegamma0v95ZOOM}
\centering
\includegraphics[width=12cm]{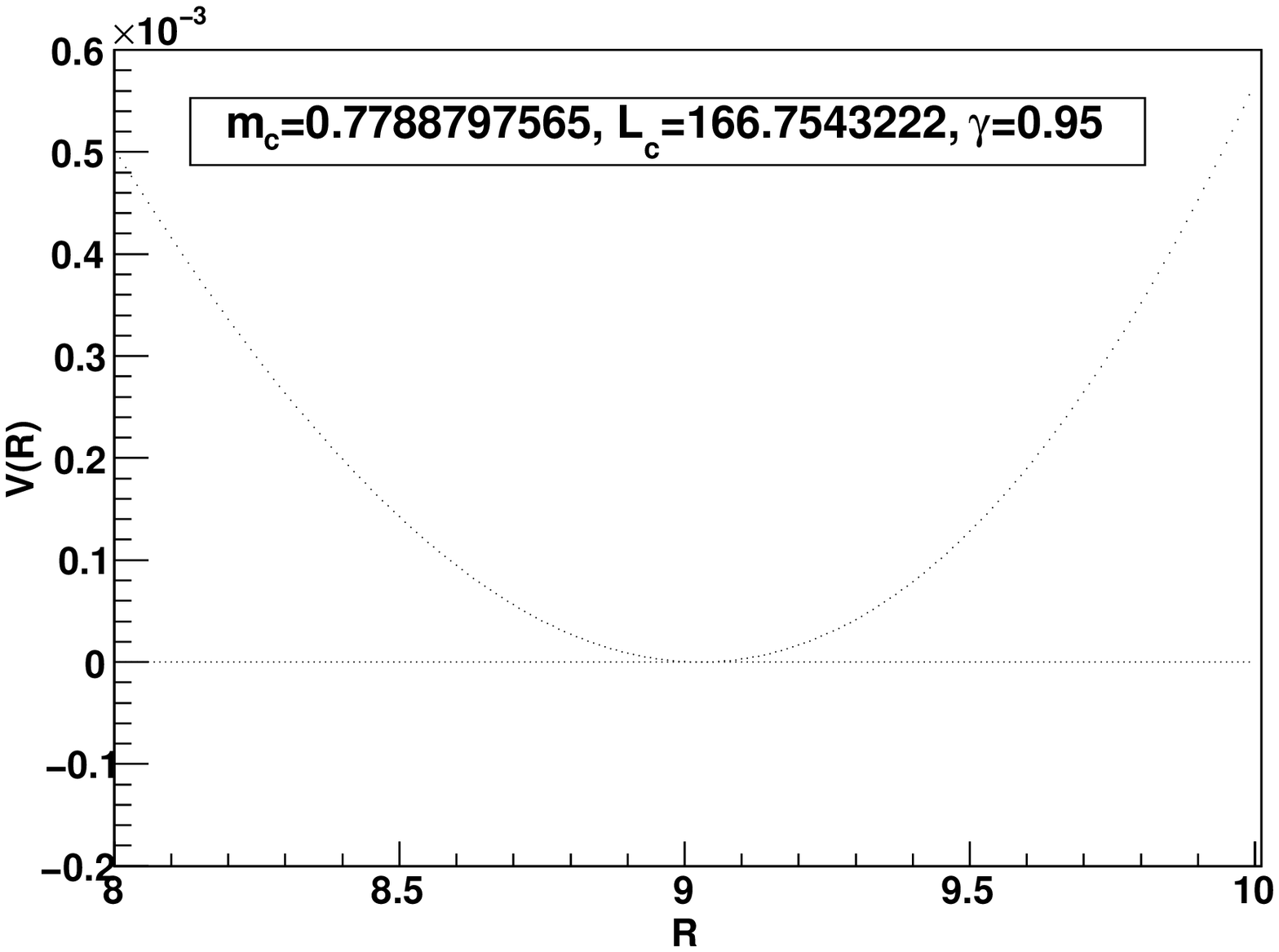}
\caption{The potential V(R) for $\gamma=0.95$ near its minimal point.}
\end{figure}

\begin{figure}
\label{generalcasegamma0v9999}
\centering
\includegraphics[width=12cm]{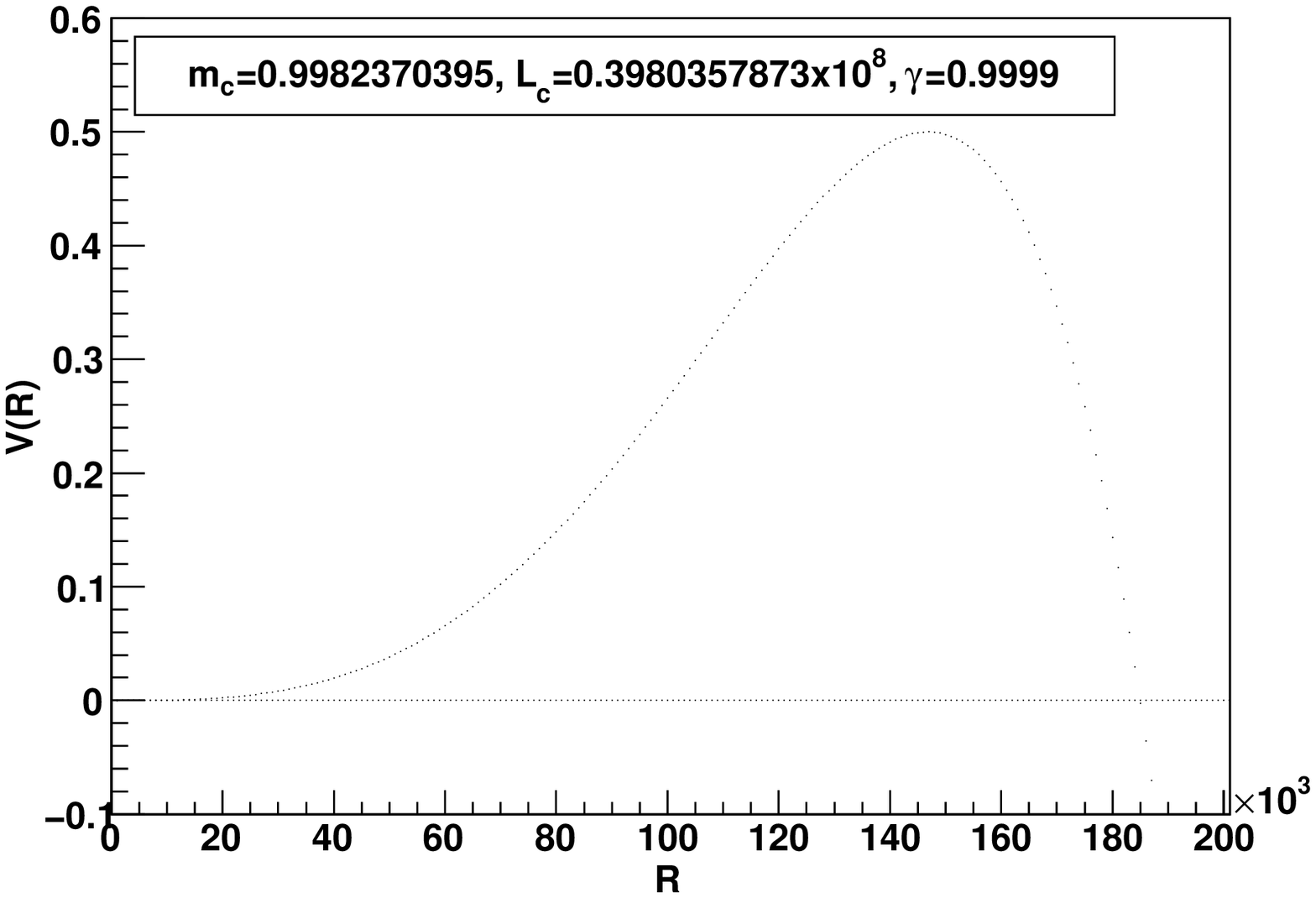}
\caption{The potential V(R) for $\gamma=0.9999$}
\end{figure}

\begin{figure}
\label{generalcasegamma0v9999ZOOM}
\centering
\includegraphics[width=12cm]{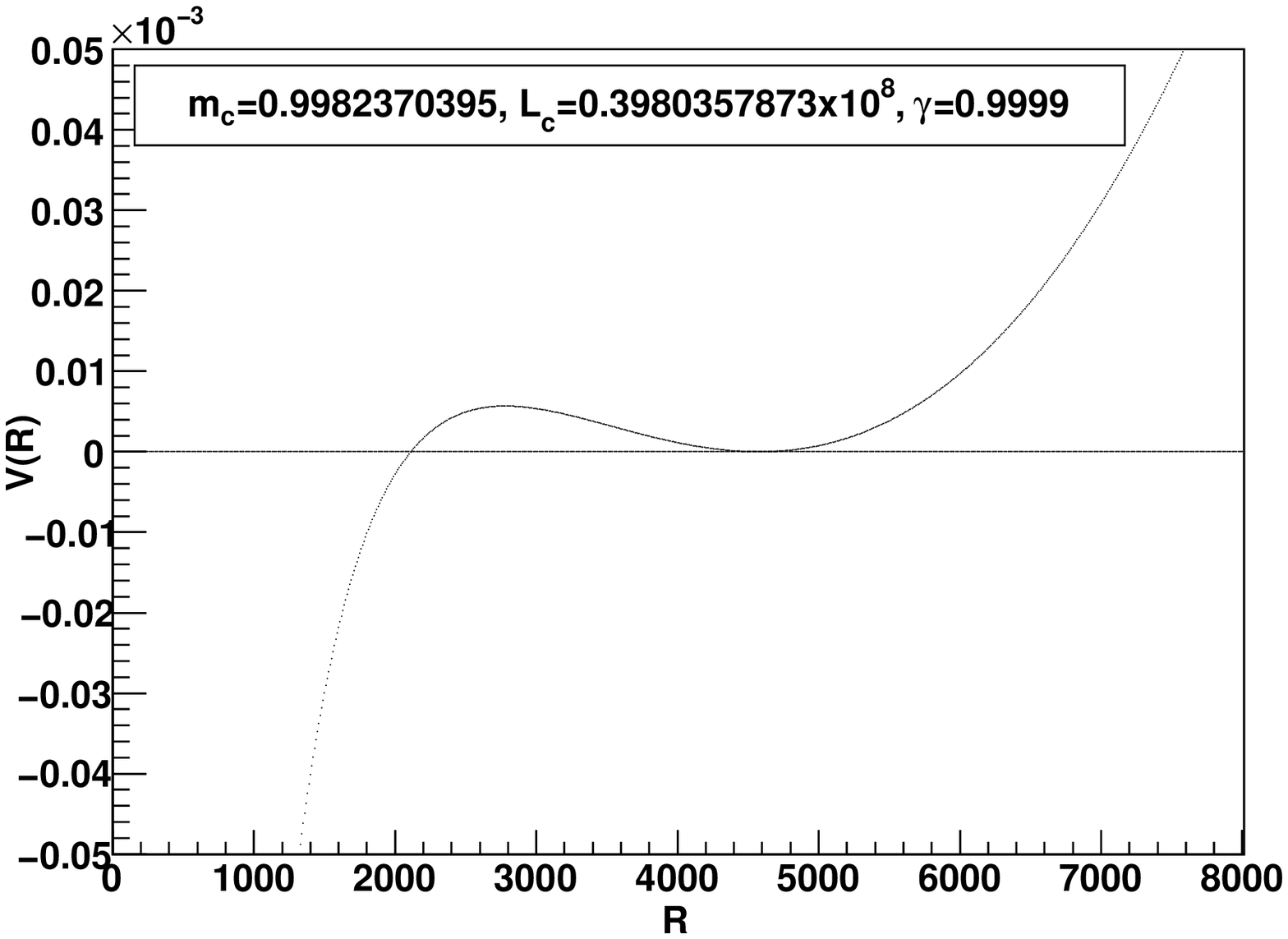}
\caption{The potential V(R) for $\gamma=0.9999$ near its minimal point.}
\end{figure}

In the cases where we can have the two types  of stable gravastars, it 
is also possible to find configurations where black holes   
are formed.  This shows  clearly that, even   gravastars indeed exist, they do not exclude
the existence of black holes.

\begin{figure}
\label{generalcasegamma1v7deltam}
\centering
\includegraphics[width=12cm]{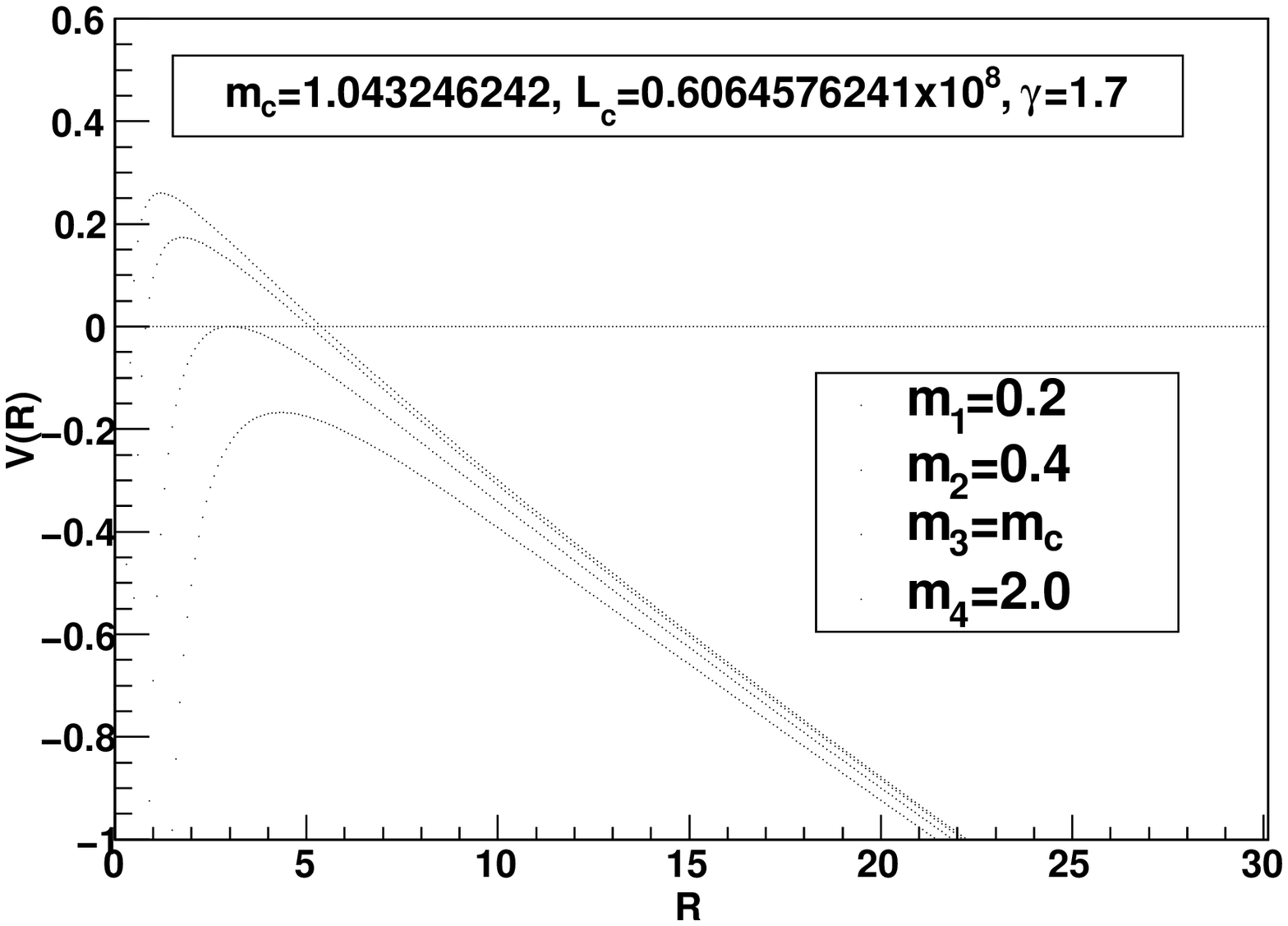}
\caption{The potential V(R) for $\gamma=1.7$. The curves from top to botton represent $m_1$ to $m_4$, respectively.}
\end{figure}

\begin{figure}
\label{generalcase1v7deltal}
\centering
\includegraphics[width=12cm]{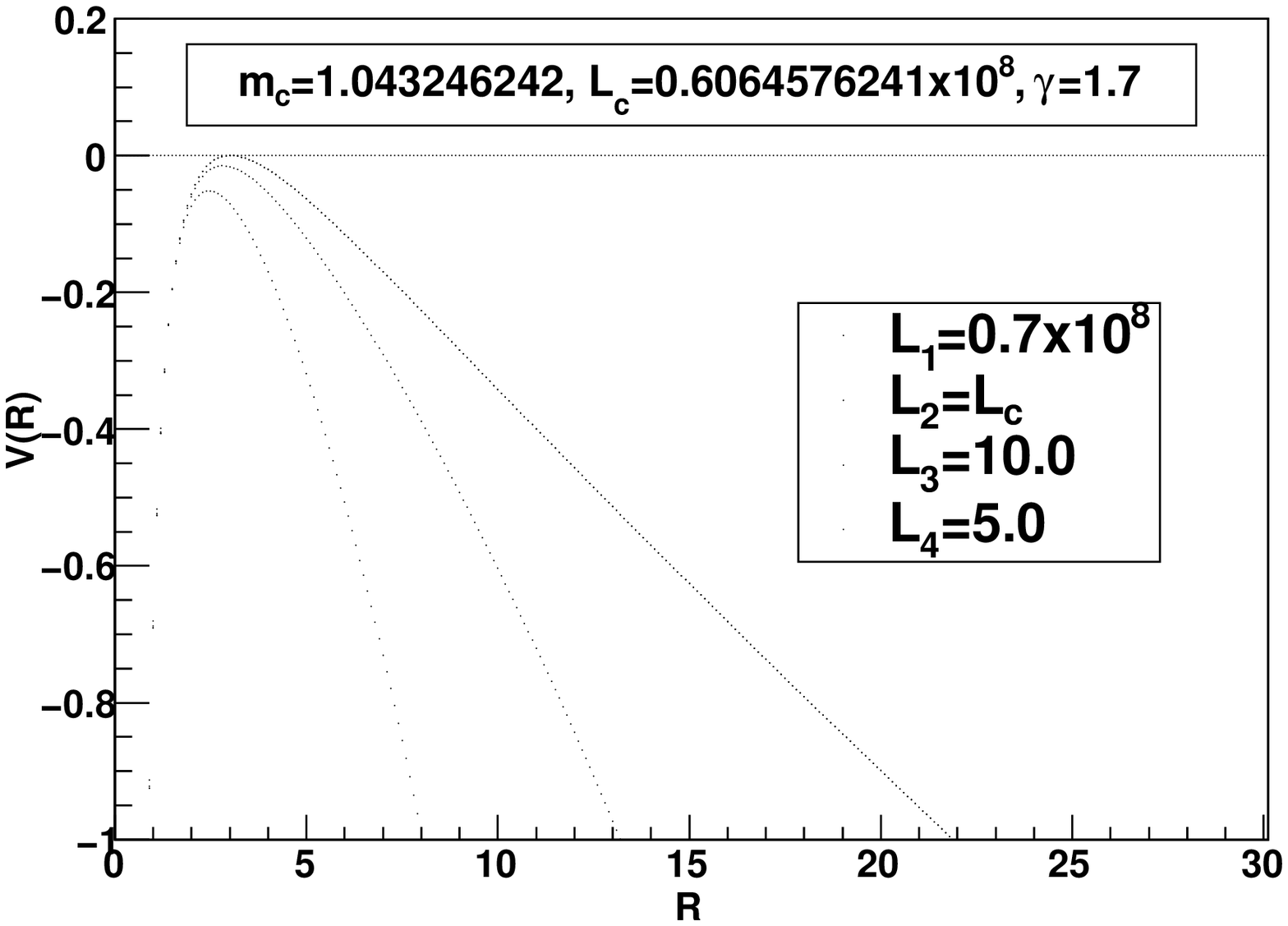}
\caption{The potential V(R) for $\gamma=1.7$. The curves from bottom to top represent $L_4$ to $L_1$, 
respectively, where the curves $L_{1}$ and $L_{2}$ coincide. Any potential curve where $L>L_{c}$ will 
coincide with the potential curve where $L=L_{c}$}
\end{figure}

\begin{figure}
\label{generalcasegamma3v0deltam}
\centering
\includegraphics[width=12cm]{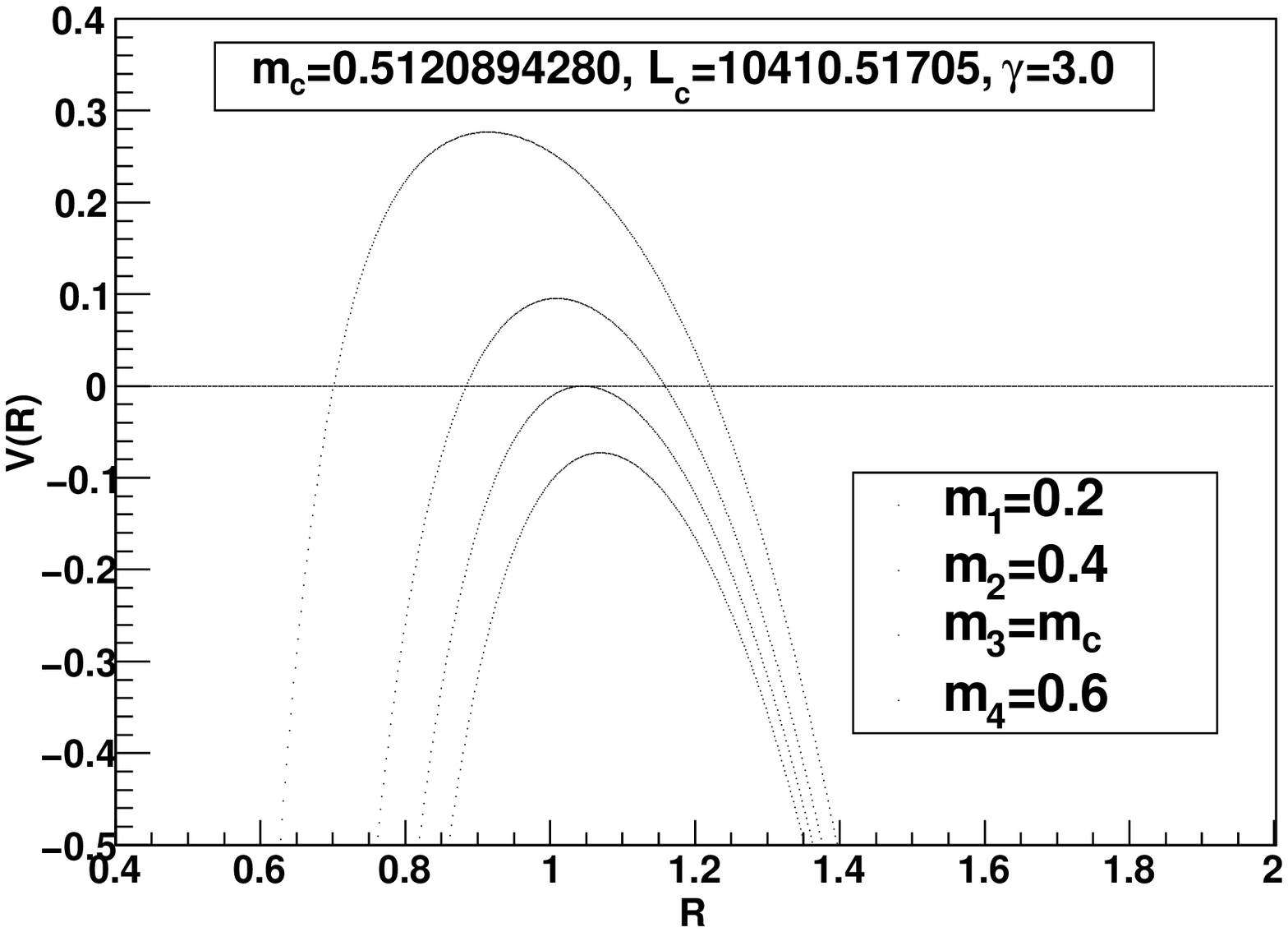}
\caption{The potential V(R) for $\gamma=3.0$. The curves from top to botton represent $m_1$ to $m_4$, respectively.}
\end{figure}

\begin{figure}
\label{generalcase3v0deltal}
\centering
\includegraphics[width=12cm]{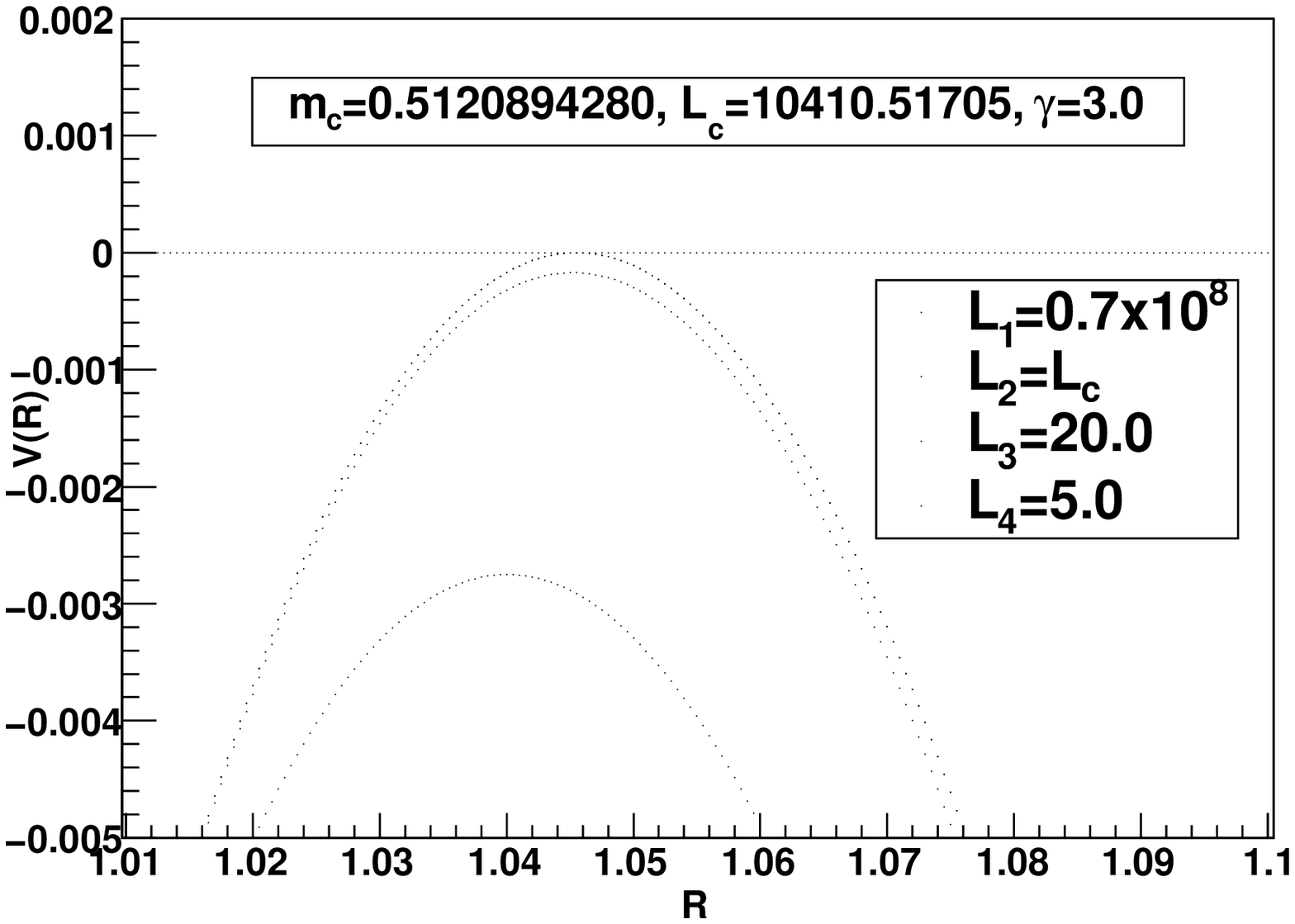}
\caption{The potential V(R) for $\gamma=3.0$. The curves from bottom to top represent $L_4$ to $L_1$, 
respectively, where the curves $L_{1}$ and $L_{2}$ coincide. Any potential curve where $L>L_{c}$ will 
coincide with the potential curve where $L=L_{c}$.}
\end{figure}

\begin{figure}
\label{generalcase5gammav0deltam}
\centering
\includegraphics[width=12cm]{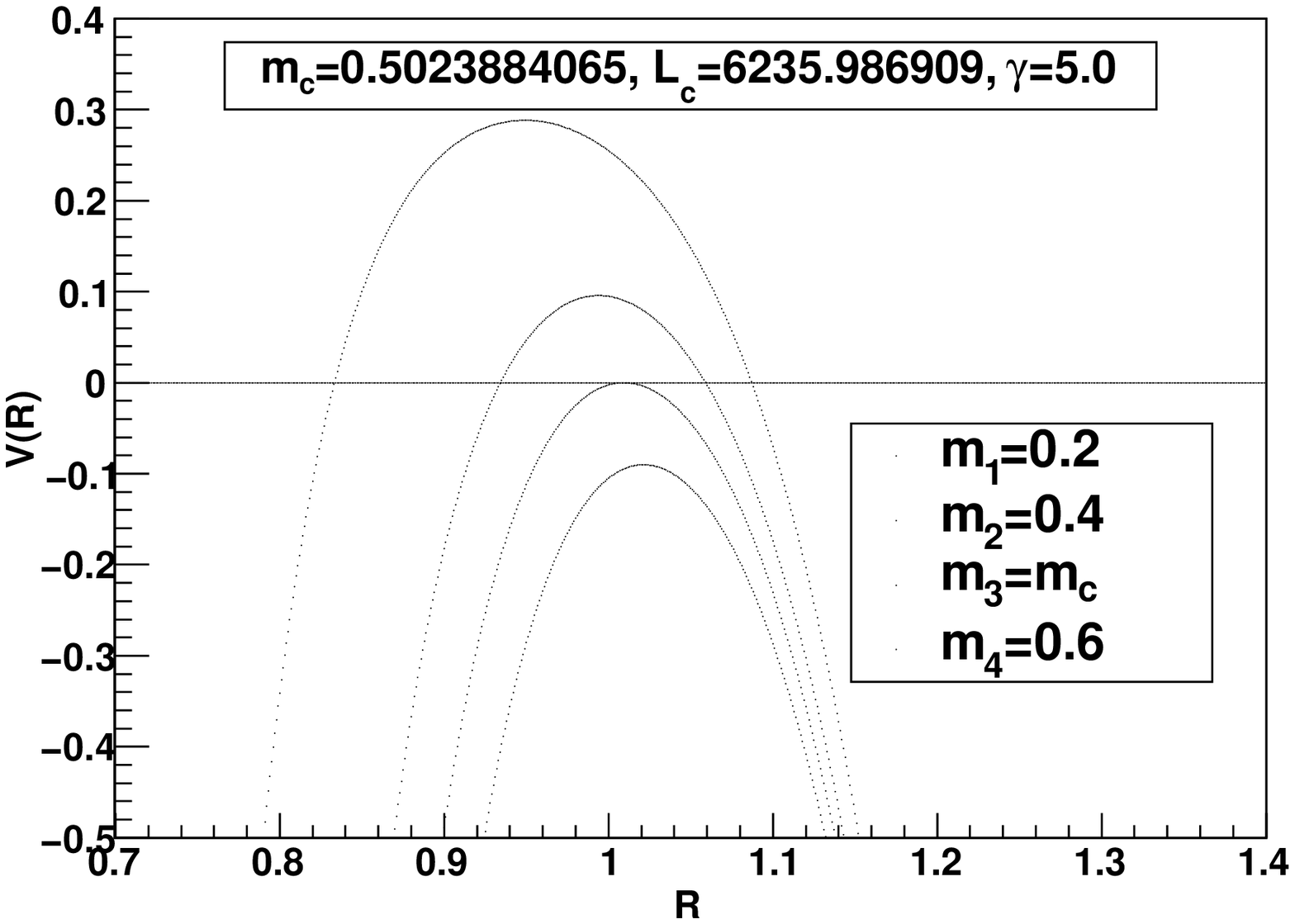}
\caption{The potential V(R) for $\gamma=5.0$. The curves from top to botton represent $m_1$ to $m_4$, respectively.}
\end{figure}

\begin{figure}
\label{generalcase5v0deltal}
\centering
\includegraphics[width=12cm]{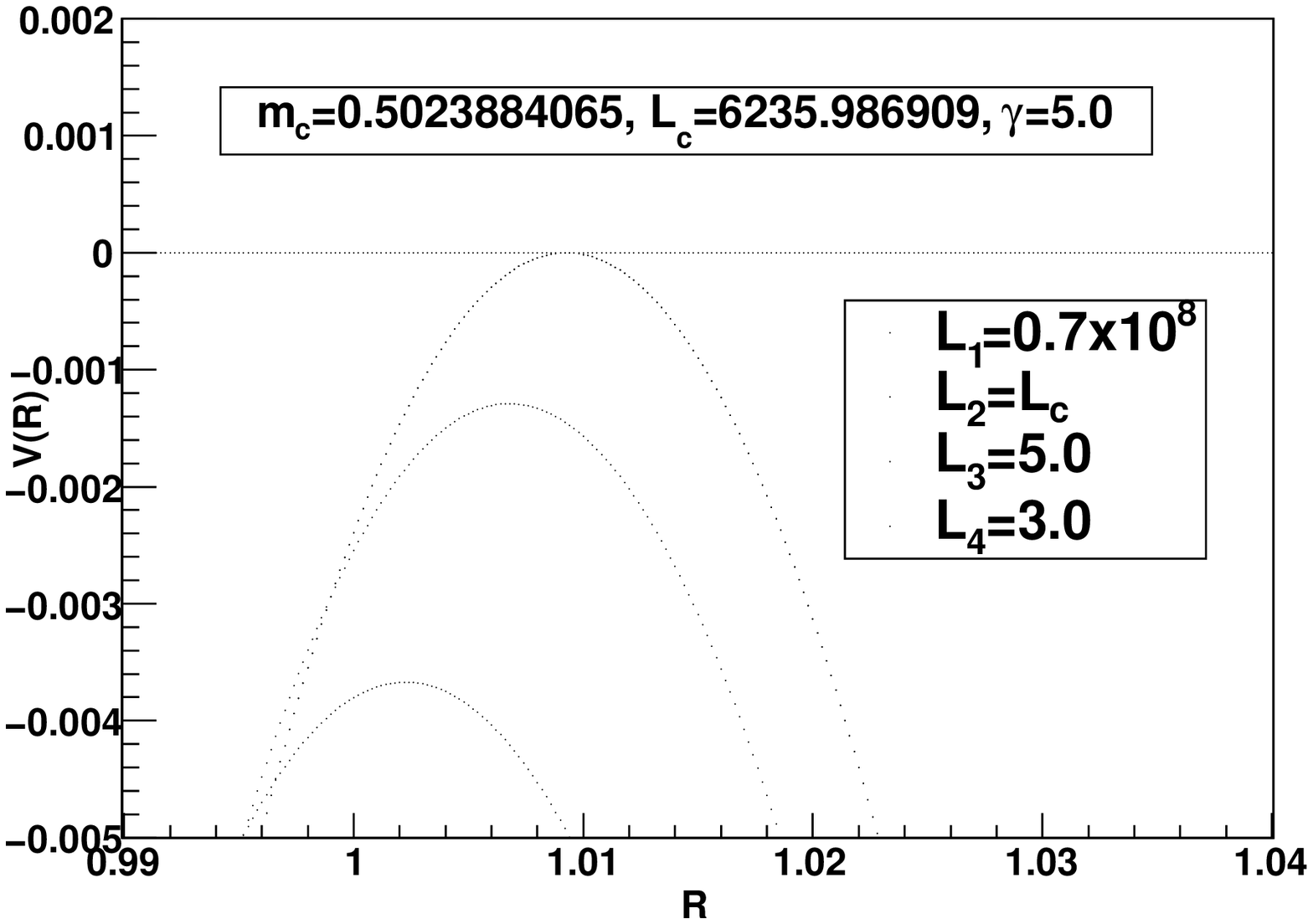}
\caption{The potential V(R) for $\gamma=5.0$. The curves from bottom to top represent $L_4$ to $L_1$, 
respectively, where the curves $L_{1}$ and $L_{2}$ coincide. Any potential curve where $L>L_{c}$ will 
coincide with the potential curve where $L=L_{c}$.}
\end{figure}

\section{Conclusions}

In this paper, we have generalized our previous work on the problem of  
stable gravastars by constructing dynamical three-layer VW models  \cite{VW04},   
which consists of an internal de Sitter space, a dynamical infinitely thin  
shell of a perfect fluid with the equation of state $p = (1-\gamma)\sigma$,
and an external Schwarzschild spacetime. 
We have shown explicitly that the final output can be a black
hole, a  ``bounded excursion"  gravastar, a stable gravastar, a Minkowski, or a de Sitter 
spacetime, depending on the total mass $m$ of the system,  the cosmological constant $\Lambda$, and  
the initial position $R_{0}$ of the dynamical thin shell. All these possibilities 
have non-zero measurements in the phase space of $m, \; \Lambda (\not= 0)$, $\gamma (< 1)$ and $R_{0}$, 
although the region  of gravastars is very small in comparison with 
that of black holes. When $\gamma \ge 1$ even with $\Lambda \not= 0$, only black holes are found.
Therefore, although the existence of gravastars cannot be 
completely excluded in these dynamical models, our results show that, even if
gravastars indeed exist, they do not exclude the existence of black holes.

\begin{acknowledgments}
The authors would like to express their gratitude to N.O. Santos for valuable
suggestions and discussions. The financial assistance from CNPq (MFAdaS, RC), 
FAPERJ (MFAdaS, RC), and FINEP (MFAdaS) is gratefully acknowledged. 
\end{acknowledgments}

\end{document}